\documentclass[twocolumn]{aastex631}
\usepackage{CJK}

\shorttitle{Orion Protostellar Disk Masses}
\shortauthors{W. Xu}

\graphicspath{{./}{figures/}}

\defcitealias{XK21b}{XK21b}
\defcitealias{Tobin2020}{T20}

\begin{document}
\begin{CJK*}{UTF8}{gbsn}

\title{Testing a New Model of Embedded Protostellar Disks Against Observation:\\
The Majority of Orion Class 0/I Disks Are Likely Warm, Massive, and Gravitationally Unstable}

\correspondingauthor{Wenrui Xu}
\email{wenruix@princeton.edu}

\author[0000-0002-9408-2857]{Wenrui Xu (许文睿)}
\affiliation{Department of Astrophysical Sciences\\
Princeton University, Peyton Hall\\
Princeton, NJ 08544, USA}

\begin{abstract}
We formulate a parametrized model of embedded protostellar disks and test its ability to estimate disk properties by fitting dust-continuum observations.
The main physical assumptions of our model are motivated by a recent theoretical study of protostellar disk formation; these assumptions include that the disk should be marginally gravitationally unstable, and that the dominant dust heating mechanism is internal accretion heating instead of external protostellar irradiation.
These assumptions allow our model to estimate reliably the disk mass even when the observed emission is optically thick and to determine self-consistently disk (dust) temperature.
Using our model to fit multi-wavelength observations of 163 disks in the VANDAM Orion survey, we find that the majority (57\%) of this sample can be fit well by our model.
Using our model, we produce new estimates of Orion protostellar disk properties. We find that these disks are generally warm and massive, with a typical star-to-disk mass ratio $M_{\rm d}/M_\star = \mathcal O(1)$ in Class 0/I.
We also discuss why our estimates differ from those in previous studies and the implications of our results on disk evolution and fragmentation.
\end{abstract}


\section{Introduction}

Accretion disks in protostellar systems\footnote{Throughout this paper, ``protostellar" system refers exclusively to a system still embedded in an infalling envelope, i.e. Class 0, Class I, and Flat Spectrum systems.} play a crucial role in the formation of stars and planets by regulating protostellar accretion and by setting the initial conditions of planet formation. However, our understanding of protostellar disks has been limited by several challenges in characterizing observationally the properties of protostellar disk populations in nearby star-forming regions and in theoretical studies of protostellar disk formation and evolution.

Our observational understanding of protostellar disk populations used to be limited mainly by the small number of observed protostellar disks (see reviews of early protostellar disk observations in \citealt{Zhao2020} and \citealt{Tobin2020}), in part due to the short timescale of protostellar evolution. The problem of sample size has been greatly alleviated by recent surveys \citep{Segura-Cox2018, Williams2019, Tobin2020}, which provide data for tens to hundreds of protostellar disks in young star-forming regions.
However, it remains challenging to estimate reliably the disk properties from these data.
In particular, it is difficult to obtain reliable estimates of disk masses from dust-continuum emission (which is what most large surveys measure).
This process involves at least two major uncertainties: whether the disk is optically thin at the observed wavelength, and the temperature of the emitting dust grains.
A common choice is to ignore these uncertainties by blindly assuming that the disk is optically thin and prescribing some arbitrary dust temperature (often $\sim$30~K). This leaves large uncertainties in the result; in particular, when the disk is optically thick, this approach can significantly underestimate the disk mass.
One can also avoid making these arbitrary assumptions and fit observation with parametrized disk models (often coupled with radiative transfer) with more degrees of freedom \citep[e.g.,][]{SheehanEisner2017, Sheehan2022}.  A major limitation of this approach, however, is that the model may be under-constrained by data. A sufficiently general model contains many ($\gtrsim$10) free parameters while currently available data sets often do not contain enough information to constrain these parameters reliably or test the physical assumptions involved in the model.

On the other hand, obtaining a clear theoretical understanding of protostellar disk formation is also very challenging.
Despite the long history of theoretical studies of this subject \citep[see a review in][]{Zhao2020}, a clear and quantitative physical picture of disk evolution is still absent. This is mainly due to theoretical difficulties in understanding the interplay between the various physical mechanisms relevant for disk formation and to technical difficulties in achieving good resolution (and numerical convergence) while modeling all relevant physical ingredients in a realistic fashion \citep{XK21a}.
However, recent studies by \citet{XK21a, XK21b} offer some optimism in resolving this problem. Using a combination of simulation and analytic theory, they argued that a typical protostellar disk should be self-regulated by gravitational instability (GI) and stay marginally unstable, and that the thermal budget of the disk is determined by a simple balance between accretion heating and radiative cooling (see a more detailed summary in Section \ref{sec:physical_assumption_Q}-\ref{sec:physical_assumption_T}). Given these constraints, the disk profile can be determined (approximately) from just a few parameters \citep[][see also Section \ref{sec:model_setup}]{XK21b}.
Yet one major caveat is that these results are based on simulations using a highly idealized initial condition (a non-turbulent pre-stellar core with uniform rotation and magnetization), and it is unclear how well the resulting physical picture can be generalized to real protostellar systems (cf. Section \ref{sec:other_formation_scenarios}).

In this paper, we try to address the aforementioned challenges by fitting observations from a recent large survey of protostellar disks, the VANDAM Orion survey \citep[][hereafter \citetalias{Tobin2020}]{Tobin2020}, using a model based on the physical picture outlined in \citet[][hereafter \citetalias{XK21b}]{XK21b}.
This will allow us to test this physical picture and obtain new (and potentially more reliable) estimates on the properties of Orion protostellar disks.

This paper is organized as follows.
We begin by describing our model in Sections \ref{sec:physical_assumptions}-\ref{sec:fit}, with a summary of our physical assumptions in Section \ref{sec:physical_assumptions}, details of the model setup in Section \ref{sec:model_setup}, and details of how we fit the model to data in Section \ref{sec:fit}.
We carefully test our model against observation in Section \ref{sec:test}, and discuss the estimates of Orion protostellar disk properties coming from our model with a focus on the physical implications of our results in Section \ref{sec:results}.
We compare our model with others in the literature to explain why we find much higher disk mass and demonstrate that our model has better predictive power in Section \ref{sec:model_comparison}. We conclude with some additional discussion in Section \ref{sec:discussion} and a brief summary in Section \ref{sec:summary}.

\section{Physical assumptions of the model}\label{sec:physical_assumptions}

\subsection{Gravitational self-regulation}\label{sec:physical_assumption_Q}

The first assumption of our model is that the disk is gravitationally self-regulated and is in a marginally gravitationally unstable state \citep[cf.][]{VB07, KratterLodato2016}.
Here we sketch out a simple argument for this assumption; for more details, see \citet{XK21a} Section 5.2 and \citetalias{XK21b} Section 4.

In an accreting protostellar system, material is accreted from the envelope\footnote{Here we use the term envelope to refer to all the infalling (and not rotationally-supported) material around the protostar-disk system. In particular, the envelope does not have to be diffuse under this definition. For example, the flattened (and relatively dense) pseudodisks that form during the collapse of quiescent, magnetized cores \citep{FiedlerMouschovias1993,GalliShu1993} are also considered as part of the envelope in this paper.} onto the disk, and then from the disk onto the protostar. During Class 0/I, the accretion onto the disk due to envelope infall happens quickly (${\sim}10^{-5}{\rm M}_\odot/{\rm yr}$ based on the typical duration of Class 0/I, \citealt{Andre2000}). Meanwhile, when the disk is gravitationally stable, the accretion rate through the disk onto the star is much lower; such accretion is mainly facilitated by angular-momentum removal from the disk via magnetic braking and magnetically-launched outflow, which are weak because of the strong ambipolar diffusion in the very weakly ionized disk (\citetalias{XK21b}; cf. \citealt{Masson2016,Zhao2018}).
Therefore, the disk mass tends to increase, eventually making it gravitationally unstable. Once the disk becomes unstable, gravitationally-excited perturbations tend to reduce the degree of instability as they can efficiently transport angular momentum outwards (which drives accretion onto the protostar) and heat up the disk.\footnote{In the literature, it is often assumed that gravitational self-regulation is mainly due to heating (``thermal saturation'', \citealt{Paczynski1978}); but in reality the decrease in surface density due to angular-momentum transport should play an equally important role, as both angular-momentum transport and heating scale with the ``effective viscosity'' of GI, or the rate at which GI extracts energy from differential rotation \citep{Gammie2001}.}
The balance between envelope infall and gravitationally-driven transport should then leave the disk in a marginally unstable state. In terms of the Toomre $Q$ parameter, this expectation translates to having
\begin{equation}
Q = \frac{\kappa \bar c_s}{\pi {\rm G}\Sigma} \approx 1{\rm -}2.\label{eq:Q}
\end{equation}
Here $\bar c_s$ the density-weighted average sound speed, $\kappa$ is the epicyclic frequency, and $\Sigma$ is the surface density. This provides a simple constraint on the surface density profile of the disk.

We stress that there are several important uncertainties related to this argument, which we summarize in Section \ref{sec:other_formation_scenarios}. Given these uncertainties, the assumption of gravitational self-regulation should not be taken for granted. We perform several tests in Section \ref{sec:test} to check whether observational data favor this assumption.

\subsection{Thermal budget and temperature profile}\label{sec:physical_assumption_T}

While gravitational self-regulation provides a constraint on the surface density profile, we also need to know the temperature profile of the disk in order to predict the dust thermal emission.

First, consider the thermal budget of the disk. We assume that the disk is mainly heated internally due to accretion, and ignore external heating due to stellar irradiation. This assumption follows the fact that the geometric thickness $H/R$ of a gravitationally self-regulated disk tends to increase towards smaller radii and the disk inner edge shields the rest of the disk from direct protostellar irradiation \citepalias[][also see the radial dependence of temperature in Appendix \ref{a:mass_scaling}]{XK21b}.
This assumption also implies that the dust temperature is approximately equal to the gas temperature; in this paper we do not distinguish between these two temperatures.

As a rough approximation, the heating rate should be comparable to the rate of gravitational energy release due to accretion (see proofs in \citealt{bgh94} and \citetalias{XK21b}). In approximate thermal equilibrium, this gives
\begin{equation}
2\sigma T_{\rm eff}^4 \sim \frac{-g_R \dot M(R)}{2\pi R} \sim \frac{-g_R \dot M}{2\pi R} = \frac{1}{2\pi}\Omega^2 \dot M.\label{eq:T_eff}
\end{equation}
Here $T_{\rm eff}$ is the effective temperature of the disk, $g_R$ is the radial gravity at the midplane, $\dot M(R)$ is the mass flux (accretion rate) through the disk at $R$, $\dot M$ is the rate of mass infall from the envelope onto the disk, and $\Omega$ is the orbital frequency. For simplicity, we assume $g_R=-{\rm G}M({<}R)/R^2$ and $\Omega = \sqrt{-g_R/R}$. 
We also assume that $\dot M(R)\sim \dot M$, which is valid when the infall rate and the disk properties all evolve slowly at a timescale comparable to the lifetime of Class 0/I.

There are two caveats regarding this assumption. First, the envelope might heat the outer disk by scattering and reemitting the radiation from the protostar-disk system \citep{DAlessio1997}. It is difficult to quantify the importance of this mechanism without introducing many free parameters as it would be sensitive to the properties of the envelope \citep{Natta1993}. Second, for a small subset of our sample the temperature in the outermost part of the disk can get below $\sim 10~{\rm K}$ and become comparable to the background temperature of interstellar irradiation; in this case this ambient irradiation could become the dominant source of heating. Generally, these limitations would lead us to underestimate the temperature in the outer disk in some cases.

Next, we want to relate $T_{\rm eff}$ to the vertical temperature profile at a given radius. This relation is determined mainly by the mechanisms of disk heating and vertical heat transport.
For the former, we assume that the heating rate per unit mass is constant (at given $R$), which is a rough but reasonable approximation for an internally heated disk.
For the latter, we assume that the vertical heat transport is dominated by radiative heat transport following \citetalias{XK21b}. While perturbations in the disk (mainly gravitationally excited spirals) could induce some forced convection, \citetalias{XK21b} demonstrated that such forced convection generally cannot be the dominant vertical heat transport mechanism and can be safely ignored \citep[cf.][]{Rafikov2007}.

Under the two assumptions discussed above, the vertical temperature profile can be well-approximated by a simple analytic form \citep[][Eq 3.11]{Hubeny1990}
\begin{equation}
T^4(\tau_{\rm R}) \approx \frac{3}{4} T_{\rm eff}^4 \left[\tau_{\rm R}\left(1-\frac{\tau_R}{2\tau_{\rm R,mid}}\right) + \frac{1}{\sqrt{3}} + \frac{1}{3\tau_{\rm P,mid}}\right].\label{eq:T_profile}
\end{equation}
Here $\tau_{\rm R}$ is the Rosseland optical depth at a given location and $\tau_{\rm R/P,mid}$ is the Rosseland/Planck optical depth at the midplane. This approximation assumes that the opacities vary slowly in depth and the heating rate per mass is approximately constant.

\subsection{Comparison with assumptions in previous studies}\label{sec:physical_assumption_comparison}

There are two main differences between our model and the assumptions in previous observational estimates of protostellar disk properties. The first is that we do not assume the emission to be optically thin at the observed wavelength.
While typical protostellar disks are likely optically thick at (and even above)  mm wavelength \citep[as suggested by their low spectral indices;][]{Li2017,Galvan-Madrid2018}, most previous mass estimates assumed optically thin emission for lack of a better method to infer the mass of obscured dust at $\tau\gtrsim 1$. Here we tackle this problem by using the physical constraints discussed above to relate the properties of the disk surface (at $\tau\lesssim 1$) to surface density and midplane temperature.

Another important difference concerns the assumed dust heating mechanism. When translating dust thermal emission to dust mass, most existing studies (e.g., \citetalias{Tobin2020}; \citealt{SheehanEisner2017,Sheehan2022}) assume that dust temperature is set mainly by protostellar irradiation, while we assume that the dust temperature is set mainly by the gas in the disk, which is heated internally due to accretion during the main accretion phase.

\section{Model setup}\label{sec:model_setup}

\subsection{Generating the disk profile}\label{sec:solve_disk_profile}

The physical assumptions given in Section \ref{sec:physical_assumption_Q} and \ref{sec:physical_assumption_T} provide enough constraints to generate the radial surface density profile and the radial and vertical temperature profile given three inputs: the mass of the protostar $M_\star$ (as the inner boundary condition), the disk size $R_{\rm d}$ (as the outer boundary), and the mass infall rate from the envelope $\dot M$. Scripts for solving the disk profile are available online at \href{https://github.com/wxu26/GIdisk2obs}{https://github.com/wxu26/GIdisk2obs}; here we sketch out a method of solution.

First we discuss how we solve the surface density and vertical temperature profile at a given radius for a given set of $\Omega, \kappa$, and $\dot M$. Here we treat ``$\sim$'' and ``$\approx$'' above as ``$=$'' when solving the disk profile and assume a constant $Q=1.5$ in our fiducial model.\footnote{Here we assume that the entire disk has constant, marginally unstable $Q$. This is a slight oversimplification. Both simulation and theory suggest that the inner part of a gravitationally self-regulated disk can be temporarily stable, and the outermost part of the disk is usually a gravitationally stable transition region \citepalias{XK21b}.}
We begin by mapping $(\tau_{\rm R,\rm mid},T_{\rm eff})$ to the vertical temperature profile and then to $(\Sigma, \bar c_s)$. For a given set of $(\tau_{\rm R,\rm mid},T_{\rm eff})$, we can solve $\tau_{\rm P, mid}$ by plugging Eq. \ref{eq:T_profile} into the constraint
\begin{equation}
\tau_{\rm P, mid} = \int_0^{\tau_{\rm R,mid}} \frac{\kappa_{\rm P}(T)}{\kappa_{\rm R}(T)}{\rm d}\tau_{\rm R}.
\end{equation}
The Rosseland and Planck mean opacities $\kappa_{\rm P},\kappa_{\rm R}$ are given in Section \ref{sec:dust}. Knowing $\tau_{\rm P, mid}$, we can then compute the vertical temperature profile $T(\tau_{\rm R})$ with Eq. \ref{eq:T_profile} and use $T(\tau_{\rm R})$ to compute $\Sigma$ (using the relation ${\rm d}\Sigma=\kappa_{\rm R}{\rm d}\tau_{\rm R}$) and $\bar c_s$.
We then (numerically) convert this map into a map from $(\Sigma/\bar c_s, T_{\rm eff})$ to other local disk properties $(\Sigma, \tau_{\rm P,mid},\tau_{\rm R,mid})$.
Since Eq. \ref{eq:Q} and \ref{eq:T_eff} directly determine $\Sigma/\bar c_s$ and $T_{\rm eff}$, we can use this map to determine the local disk properties.

To generate a disk profile, we just need to update the radial profile of $(\Omega, \kappa)$ iteratively using the $\Sigma$ profile calculated with current estimates of $(\Omega, \kappa)$ until the results converge. Here $M_\star$ and $R_{\rm d}$ are used as boundary conditions for this process.

\subsection{Generating mock observation}

Using the disk temperature profile $T(R,\tau_{\rm R})$ generated from our model, we can produce mock observations at a given wavelength (frequency).

We first compute the intensity $I_\nu(R)$ at disk surface using the temperature profile $T(R,\tau_{\rm R})$. Since the disk is often optically thick and the scattering opacity is generally higher than absorption opacity at the observed wavelengths, we need to include scattering in our calculation \citep{Zhu2019}. 
For simplicity, we assume that the disk is geometrically thin; this reduces solving $I_\nu$ at given $R$ to a 1D problem. The intensity is given by
\begin{equation}
I_\nu = \int_0^{\tau_{\nu,\rm tot}} e^{-\mu\tau_\nu} S_\nu \mu {\rm d}\tau_\nu.
\end{equation}
Here $\tau_\nu$ is the optical depth from disk surface (along vertical direction), $\tau_{\nu, \rm tot}$ is the total optical depth, $\mu=\cos I$ with $I$ being the inclination of the disk, and $S_\nu$ is the source function. We map $\tau_{\rm R}$, which is used to specify the temperature profile, to $\tau_\nu$ using the relation
\begin{equation}
\kappa_{\rm R}^{-1} {\rm d}\tau_R = (\kappa_{\nu, \rm abs} + \kappa_{\nu, \rm sca}^{\rm eff})^{-1}{\rm d}\tau_\nu.
\end{equation}
Here $\kappa_{\nu, \rm abs}$ is the absorption opacity and $\kappa_{\nu, \rm sca}^{\rm eff} = (1-g_\nu)\kappa_{\nu, \rm sca}$ is the effective scattering opacity, where $g_\nu$ is the forward-scattering parameter and the factor $(1-g_\nu)$ accounts for anisotropic scattering \citep{Ishimaru1978}.
The source function $S_\nu$ is given by
\begin{equation}
S_\nu = (1-\omega_\nu)B_\nu(T) + \omega_\nu J_\nu(\tau_\nu),
\end{equation}
where $B_\nu(T)$ is the Planck function, $J_\nu = \frac{1}{4\pi}\int I_\nu{\rm d}\Omega$ is the isotropic intensity, and $\omega_\nu = \kappa_{\nu, \rm sca}^{\rm eff}/(\kappa_{\nu, \rm abs} + \kappa_{\nu, \rm sca}^{\rm eff})$ is the single-scattering albedo.
Now we only need to solve for $J_\nu$. Under the Eddington approximation, the second moment of the radiative transfer equation becomes
\begin{equation}
\frac 13 \frac{{\rm d}^2 J_\nu}{{\rm d}\tau_\nu^2} = (1-\omega_\nu) (J_\nu-B_\nu),
\end{equation}
and we solve it (numerically) with the boundary condition of no incoming radiation field, which under the two-stream
approximation is given by ${\rm d}J_\nu/{\rm d}\tau_\nu = \pm\sqrt{3} J_\nu$ at top/bottom surface \citep{MiyakeNakagawa1993}. In the limit of high optical depth, including scattering generally reduces $I_\nu$ by a factor of $\mathcal O(\sqrt{1-\omega_\nu})$ \citep{Zhu2019}.

We then use the $I_\nu(R)$ profile to generate a mock observation image by orienting the disk using the position angle (PA) and inclination estimates from \citetalias{Tobin2020}, which are based on the deconvolved shape of the best-fit gaussian profile of the 0.87 mm image, and convolve the image with a 2D gaussian beam whose widths and orientation are identical to the synthesized beam of the corresponding observation. This produces the intensity profile $I_\nu$ in Jy per beam.

%
%
\subsection{Dust opacity and disk truncation}\label{sec:dust}

\begin{figure}
    \centering
    \includegraphics[scale=0.66]{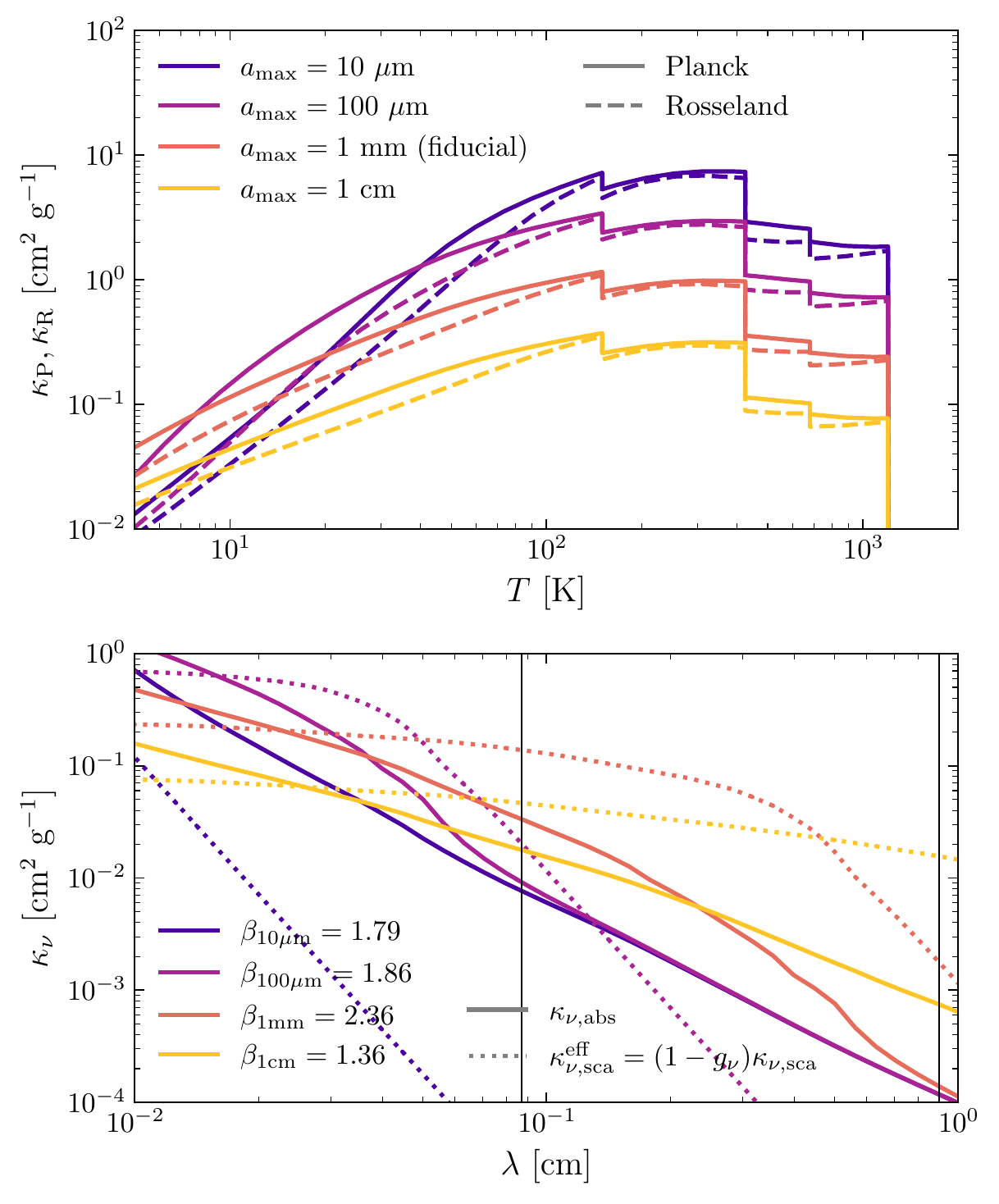}
    \caption{Top panel: Planck and Rosseland mean opacities $\kappa_{\rm P},\kappa_{\rm R}$ for dust models with different maximum grain size $a_{\rm max}$. (For $a_{\rm max}<10~\mu$m, the results are nearly identical to $a_{\rm max}=10~\mu$m.) Bottom panel: absorption opacity $\kappa_{\nu,\rm abs}$ and effective scattering opacity $\kappa_{\nu,\rm sca}^{\rm eff}$ before any dust sublimation ($T<150$ K). We also label the dust opacity indices $\beta$ between 0.87~mm and 9~mm (vertical black lines) for $\kappa_{\nu,\rm abs}$.}
    \label{fig:opacity}
\end{figure}

Generating the disk profile and mock observation both require knowledge on the opacity of the disk, which is dominated by dust opacity. Here we assume a constant dust-to-gas mass ratio of 0.01 and a power-law grain-size distribution with ${\rm d}n/{\rm d}a\propto a^{-3.5}$ \citep{Mathis1977} having a maximum grain size $a_{\rm max}=1$~mm and a minimum grain size $a_{\rm min}\to 0$. (As long as the minimum grain size $a_{\rm min}\ll a_{\rm max}$, the opacities are insensitive to the exact choice of $a_{\rm min}$.) The assumption of a constant dust-to-gas ratio should be reasonable, as the dust grains are well coupled to the gas (Appendix \ref{a:grain_coupling}).

Our choice of $a_{\rm max}$ assumes that the dust-size distribution is more similar to that in protoplanetary disks than that in the ISM, which would have $a_{\rm max}\lesssim \mu$m. This is because grain growth is expected to proceed quickly in the disk before approaching the equilibrium between coagulation and fragmentation \citep{Birnstiel2011}. Still, the exact value of $a_{\rm max}$ remains highly uncertain, as $a_{\rm max}$ can vary significantly across different disks and vary radially within a disk. To address this uncertainty, while we use a fiducial value of $a_{\rm max}=1$ mm, we also produce models with $a_{\rm max} = 10~\mu$m, 100~$\mu$m, and 1~cm and compare the results across these models.

The dust opacities are computed using the DSHARP opacity package \citep{Birnstiel2018}. The assumptions for grain porosity and composition also follows that of \citet{Birnstiel2018}, with zero-porosity grains composed of 20\% water ice, 40\% refractory organics, 7\% troilite, and 33\% silicates by mass. Since the temperatures in the inner part of our model profiles are often several hundred K and above, we need to include the effect of dust sublimation. We adopt the sublimation temperatures from \citet{Pollack1994} and remove water ice, refractory organics, troilite, and silicates from our dust mixture (while adjusting the dust-to-gas ratio accordingly) at 150, 425, 680, and 1200~K, respectively. Here for simplicity we have dropped the density dependence of water ice and silicate sublimation temperatures. Beyond 1200~K, we assume that $\kappa_{\rm P},\kappa_{\rm R}\to 0$ as the remaining gas opacity there is much lower than the dust opacity below 1200~K. In Fig. \ref{fig:opacity} we summarize the opacities for dust models with different $a_{\rm max}$.

Once disk material exceeds 1200~K, it can barely cool by radiation due to the drop in opacity and thermal equilibrium can only be achieved at a much higher temperature when gas opacity becomes sufficiently high. This results in a steep increase of temperature at $R<R_{1200\rm K}$, where $R_{1200\rm K}$ is the radius where midplane temperature first reaches 1200~K.
The high temperature at $R<R_{1200\rm K}$ would lead to well-coupled magnetic field because the thermal ionization of potassium increases disk ionization exponentially above $\sim 1000$~K \citep{Umebayashi1983}.
In this regime, our model is no longer applicable, and the disk is likely to be regulated by MRI and magnetized outflow which efficiently transport angular momentum (with effective $\alpha \gtrsim 0.01$) to keep the surface density at $R\lesssim R_{1200\rm K}$ low \citep[cf.][]{Gammie1996}.
For simplicity, we excise this region from our calculation and assume that the surface density and flux density are zero at $R<R_{1200\rm K}$.

\subsection{Choice of model parameters}\label{sec:parameter}

\begin{table}
\caption{Summary of fiducial model parameters\label{tab:parameter}}
\begin{tabular}{p{10em}p{7.5em}}
\tableline
\multicolumn{2}{l}{\bf Free parameters}\\
\multicolumn{2}{l}{$M_\star, R_{\rm d}$}\\
\tableline
\multicolumn{2}{l}{\bf Fixed parameters}\\
Toomre $Q$ & 1.5 \\
Max grain size $a_{\rm max}$ & 1 mm\\
$\dot M/M_\star$ & $10^{-5}~{\rm yr}^{-1}$\\
dust-to-gas ratio & 0.01\\
grain porosity & 0\\
\tableline
\end{tabular}
\end{table}

We conclude this section by summarizing all parameters of our model and discussing the choice of free (and fixed) parameters. The disk profile can be solved with only three parameters: the mass of the star $M_\star$, the disk size $R_{\rm d}$, and the rate of mass infall (accretion) from envelope $\dot M$ (Section \ref{sec:solve_disk_profile}). One intuitive choice is to leave all three parameters as free parameters; this, however, will be problematic for unresolved disks, where observation only has two degrees of freedom corresponding to the integrated flux at the two observed wavelengths. (There is a little additional information as a finite disk size will always cause the profile to deviate from being exactly Gaussian; but such information is negligible when the disk diameter is significantly smaller than the beam size.) Therefore, in order to fit all observed systems, we want our model to have at most two free parameters. This is achieved by assuming a constant $\dot M/M_\star = 10^{-5}{\rm ~yr}^{-1}.$
The choice of this ratio is motivated by the typical lifetime of the main accretion phase (Class 0 and Class I). The mass dependence is meant to capture (to some extent) the possible difference in accretion rates between low-mass and high-mass systems, but not the variation of accretion rate in a single system as the protostar gains mass. This is of course a very crude approximation; it is unclear whether the duration of the main accretion phase has any mass dependence, and it does not correctly capture the temporal variation of accretion rate within the main accretion phase (where $\dot M$ should remain approximately constant or decrease as $M_\star$ and the age of the system increase). We address the large systematic uncertainty in our assumed $\dot M$ by quantifying how a different choice of $\dot M/M_\star$ impacts the result in Section \ref{sec:systematic_uncertainties} and \ref{sec:test_agreement}.

We also comment that there are several other ways for estimating $\dot M$, but they each have their own limitations. First, one can simply choose a fixed $\dot M$, but it would not be reasonable to assume that all systems -- which span several orders of magnitude in mass -- have identical accretion rates. Besides, we find that the best-fit model for some low-mass systems would have $M_\star\to 0$ if we assume a fixed $\dot M\sim 10^{-5}~{\rm M}_\odot~{\rm yr}^{-1}$. Another possibility is to estimate $\dot M$ using the luminosity of the protostar $L_\star$. However, $L_\star$ depends on the accretion rate onto the star $\dot M_\star$ and can be highly variable (potentially due to the modulation of the disk), such that an instantaneous estimate of $\dot M_\star$ is not necessarily a good approximation of the mass infall rate $\dot M$ \citep[cf.][]{OffnerMcKee2011,Zakri2022}. Therefore, our choice of a constant $\dot M/M_\star$ is probably already the best we can do before observations reach better resolution (allowing us to fit $\dot M$ as a free parameter) or provide direct estimates of $\dot M$ (e.g., through envelope dynamics, which is already available for a small number of systems, e.g., \citealt{Kristensen2012,Pineda2012}).

In addition to our two free parameters $M_\star, R_{\rm d}$ and our fixed $\dot M/M_\star$, there are a few additional fixed parameters to the model, including an assumed constant Toomre $Q$ of 1.5 (following the idea of gravitational self-regulation) and parameters for the fiducial dust model (Section \ref{sec:dust}). We summarize the fiducial model parameter choices in Table \ref{tab:parameter}. In Sections \ref{sec:systematic_uncertainties}, \ref{sec:test_agreement}, and Appendix \ref{a:uncertainty} we also discuss how different choices of these fixed parameters impact the agreement between model and observation and the estimated disk properties.

\section{Fitting the model to observation}\label{sec:fit}

\subsection{Sample selection}

We use observations from the VANDAM Orion survey \citep[\citetalias{Tobin2020};][]{VANDAMOrionALMA, VANDAMOrionVLA}, a large survey of protostellar systems in the Orion molecular clouds with ALMA (0.87~mm) and VLA (9~mm) at ${\sim}40$~au resolution.
Our sample consists of all systems that are detected at both wavelengths and have positive deconvolved major and minor axes at 0.87~mm (which are used for estimating inclination).
This gives a total of 163 systems; among them, 98 are Class 0, 40 are Class I, 21 are Flat Spectrum, and 4 are unclassified.

\subsection{Fitting}

For each system, we vary the two free parameters of the model, $M_\star$ and $R_{\rm d}$, to minimize the error between observed and model images (flux density). We fit the model with images as opposed to visibilities (in frequency space) because it is easier to characterize the systematic uncertainty of the model (due to the oversimplifications we made) in real space rather than in frequency space.

We evaluate the error between the observed flux density $I_{\rm obs}$ and the model flux density $I_{\rm model}$ using
\begin{equation}
\chi^2 \equiv \min\left\{ \frac{| I_{\rm obs}-I_{\rm model}|^2}{2\sigma_{\rm obs}^2}, \frac{\log(I_{\rm obs}/I_{\rm model})^2}{2\sigma_{\rm log~model}^2}\right\}.
\end{equation}
Here $\sigma_{\rm obs}$ is the observation uncertainty, estimated with the RMS flux density of each field of observation, and $\sigma_{\rm log~model}$ captures the systematic uncertainty of the model due to the many oversimplified assumptions in our model. The errors in disk properties due to these oversimplifications are generally of order unity (cf. \citetalias{XK21b} Fig. 16); therefore we choose $\sigma_{\rm log~model} = \log(2)/2$, such that $\pm 1\sigma$ covers a factor of 2 difference in $I_{\rm obs}/I_{\rm model}$. $\chi^2$ roughly corresponds to the minus log likelihood (per beam) of observing the given deviation between model and observation.

In order to find the best-fit model, we minimize
\begin{equation}
l \equiv \int \chi_A^2 S_{B,A}^{-1} {\rm d}S  + \int \chi_V^2 S_{B,V}^{-1} {\rm d}S.
\end{equation}
Here $S_{B, A/V} = \frac{\pi}{4\ln 2}b_{A/V, \rm maj}b_{A/V, \rm min}$ is the beam area (converted to au$^2$ to match ${\rm d}S$), with $b$ being the FWHM. Roughly speaking, $\int ...~ S_{B,A/V}^{-1} {\rm d}S$ represents a summation over beams.
The integral covers a square region around the protostar with width 800 au or $4{\times}$ the disk radius estimate in \citetalias{Tobin2020}, whichever is larger. $l$ can be interpreted as the minus log likelihood of observing the given error. 


We minimize $l$ under the constraint that the model flux density at 0.87 mm (before blurring) needs to be ${\geq} \sigma_{\rm obs}$ at the disk edge. This is because the fit becomes much less reliable when the flux is below detection limit, as the model tends to fit the envelope emission around the disk (which is not included in our model) by incorrectly increasing the disk size. When the fit is affected by this constraint, the best-fit disk size should be interpreted as an estimate for the lower limit of the true disk size. For our fiducial model, 31 systems are in this regime.

\subsection{Quantifying systematic uncertainties}\label{sec:systematic_uncertainties}

In order to avoid overfitting, we have reduced the number of free parameters in our model by making a series of oversimplifying and/or arbitrary assumptions. It is important to evaluate whether our estimated disk properties are sensitive to these assumptions and quantify the systematic uncertainties in our results.
This is done by fitting our sample with different assumed model parameters, and using the measured dependence to estimate the systematic uncertainties in our results. The details of this process are documented in Appendix \ref{a:uncertainty}. In summary, assuming one order-of-magnitude uncertainty in $a_{\rm max}$ and $\dot M/M_\star$ and an order-unity uncertainty in $Q$, the uncertainties in key disk properties are generally of order unity and at most a factor of $\sim 3$ (see Table \ref{tab:uncertainty}).

\section{Testing the model against observation}\label{sec:test}

\begin{figure*}
    \centering
    \includegraphics[scale=.6]{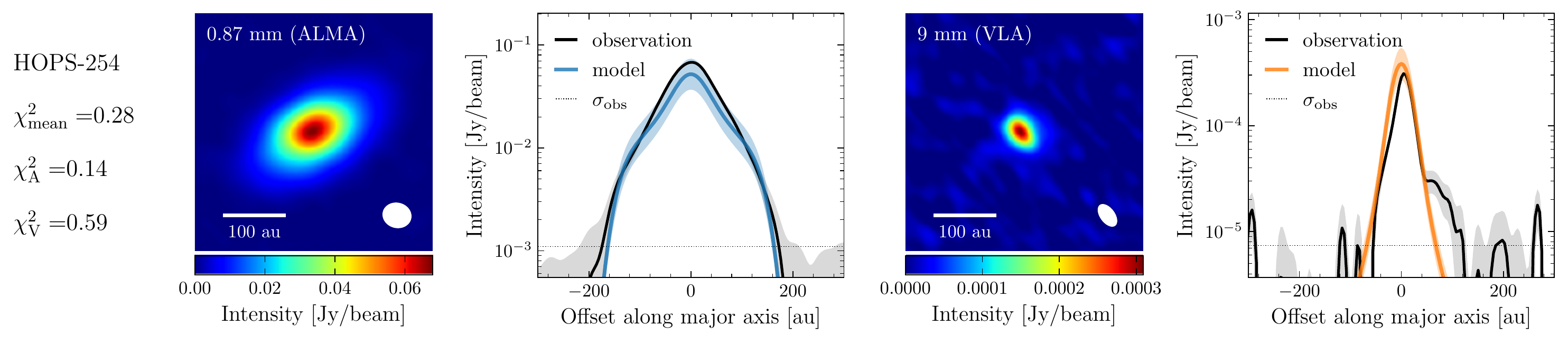}
    \includegraphics[scale=.6]{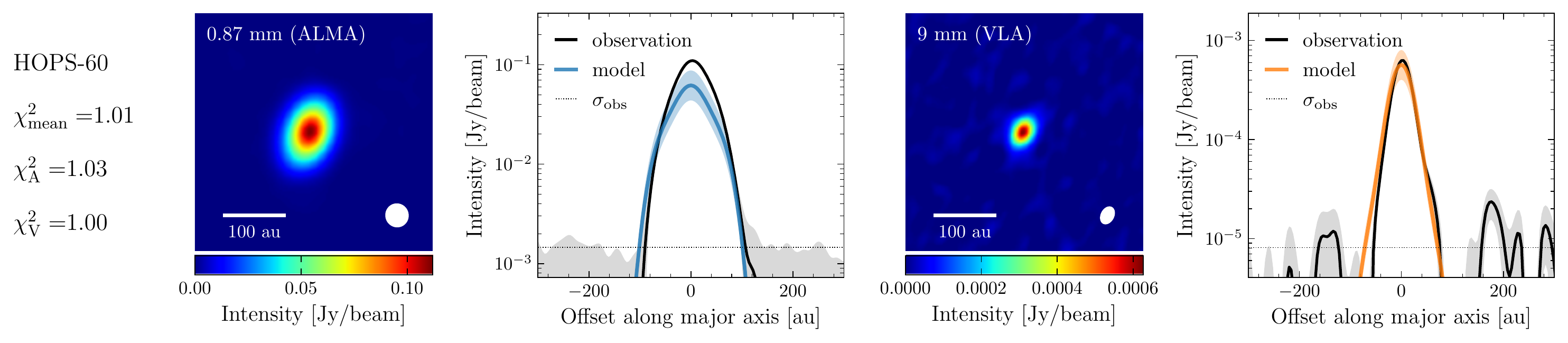}
    \includegraphics[scale=.6]{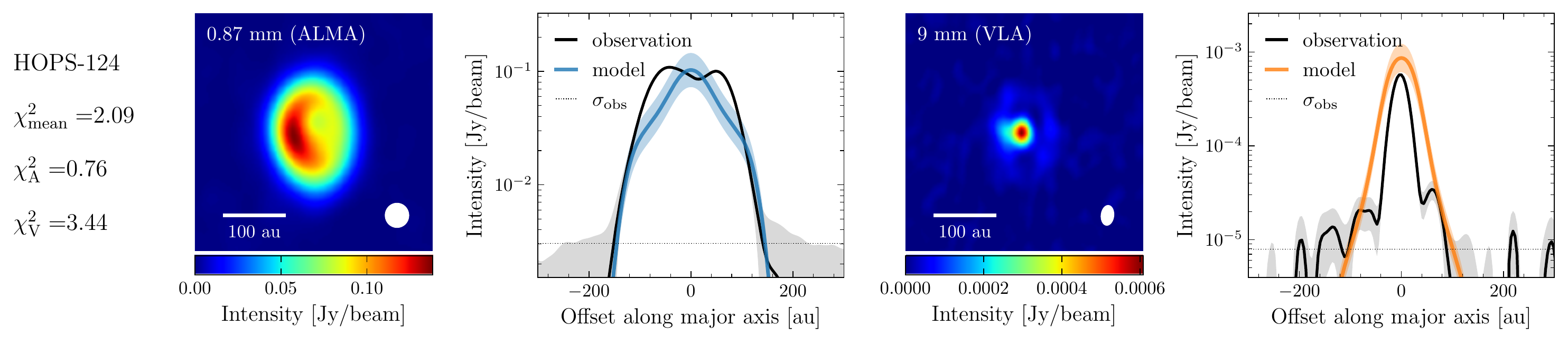}
    \caption{Examples of well-resolved systems with different $\chi_{\rm mean}^2$. Each row shows the observations at both wavelengths and compares the best-fit model with observation by taking cuts along the disk major axis (positive offset is in the direction of increasing RA). Shaded areas show $1\sigma$ observation and model uncertainties. The first two panels show examples where the model agrees well with the observations ($\chi_{\rm mean}^2\lesssim 1$); $57\%$ of our sample are in this regime. The last panel shows an example where the model cannot fit the observation very well. For this particular system this is due to the presence of substructure (which at current resolution is uncommon in the sample). In general, the lack of a good fit could be due to a variety of reasons, such as disk being inconsistent with our model, envelope contamination, and binarity.}
    \label{fig:compare_fit_with_img}
\end{figure*}

\begin{figure}
    \centering
    \includegraphics[scale=0.66]{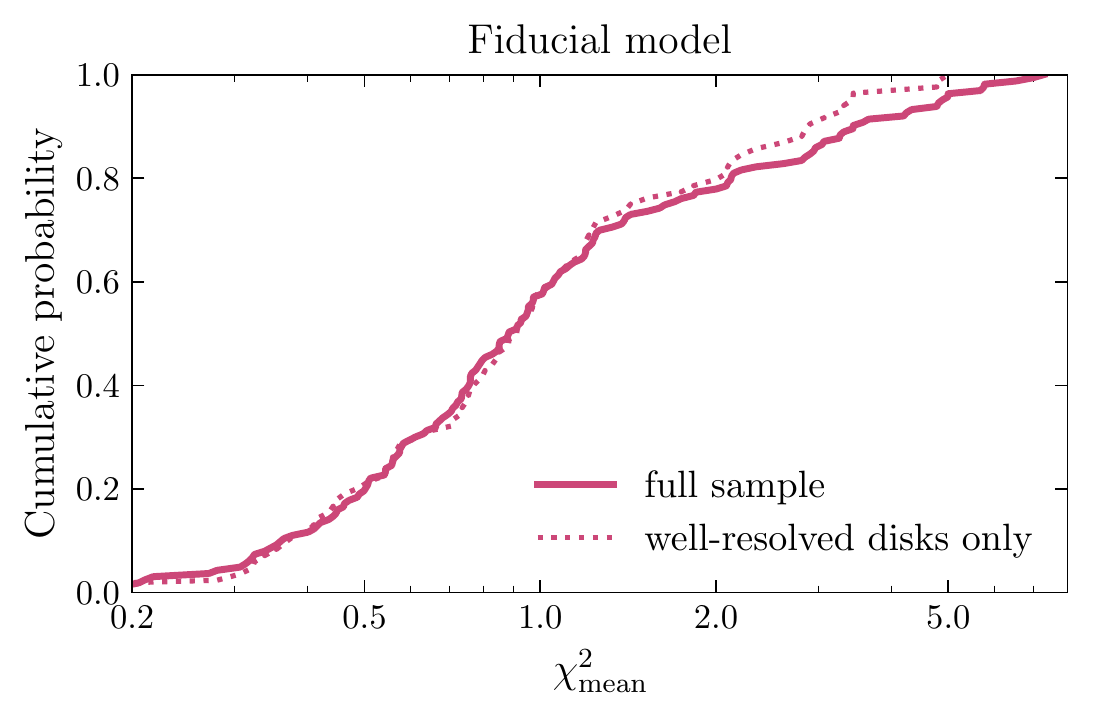}
    \caption{Cumulative distribution of model error $\chi_{\rm mean}^2$ for our fiducial model. $57\%$ of our sample can be fit well with our model ($\chi_{\rm mean}^2\leq 1$). The result is similar if we limit the sample to only well-resolved disks (\citetalias{Tobin2020} disk size estimate ${>}50$~au).}
    \label{fig:chi_sq_fid}
\end{figure}

\begin{figure}
    \centering
    \includegraphics[scale=0.66]{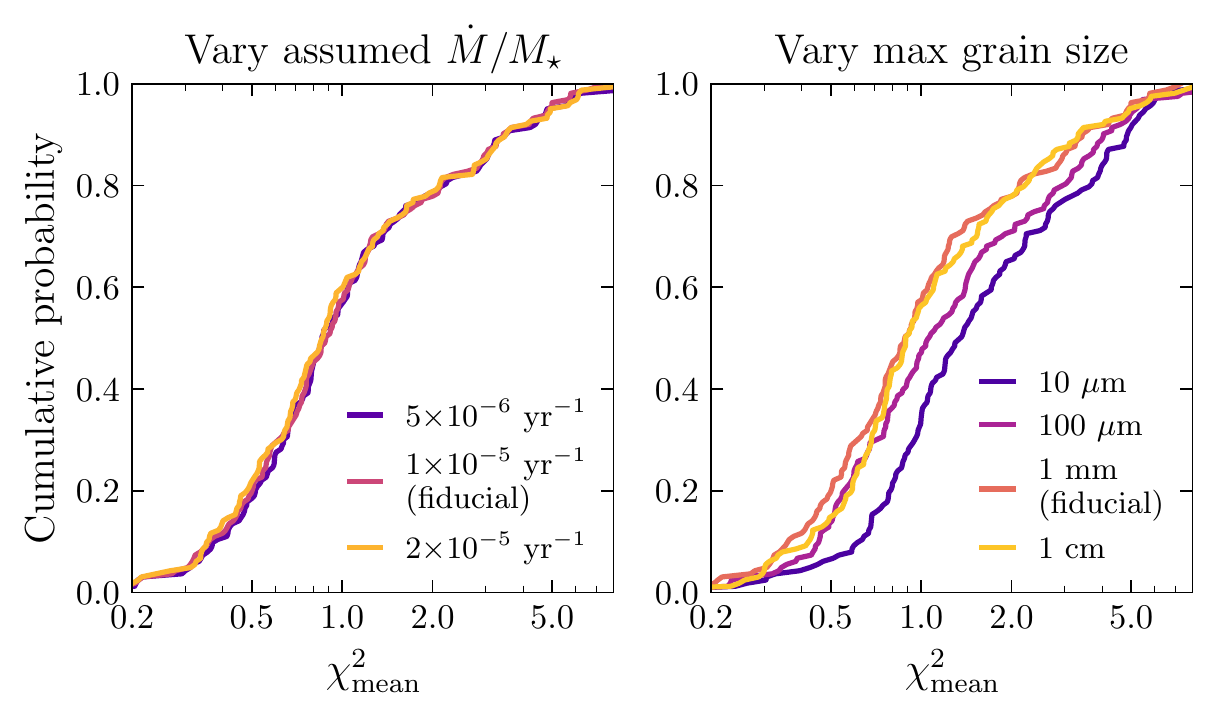}
    \caption{Cumulative distribution of model error $\chi_{\rm mean}^2$ when we vary the assumed $\dot M/M_\star$ (left panel) and the assumed dust size distribution (right panel). The agreement remains overall similar except for very small grain size ($a_{\rm max}=10\mu{\rm m}$).}
    \label{fig:chi_sq_Mdot_and_amax}
\end{figure}

\begin{figure}
    \centering
    \includegraphics[scale=0.66]{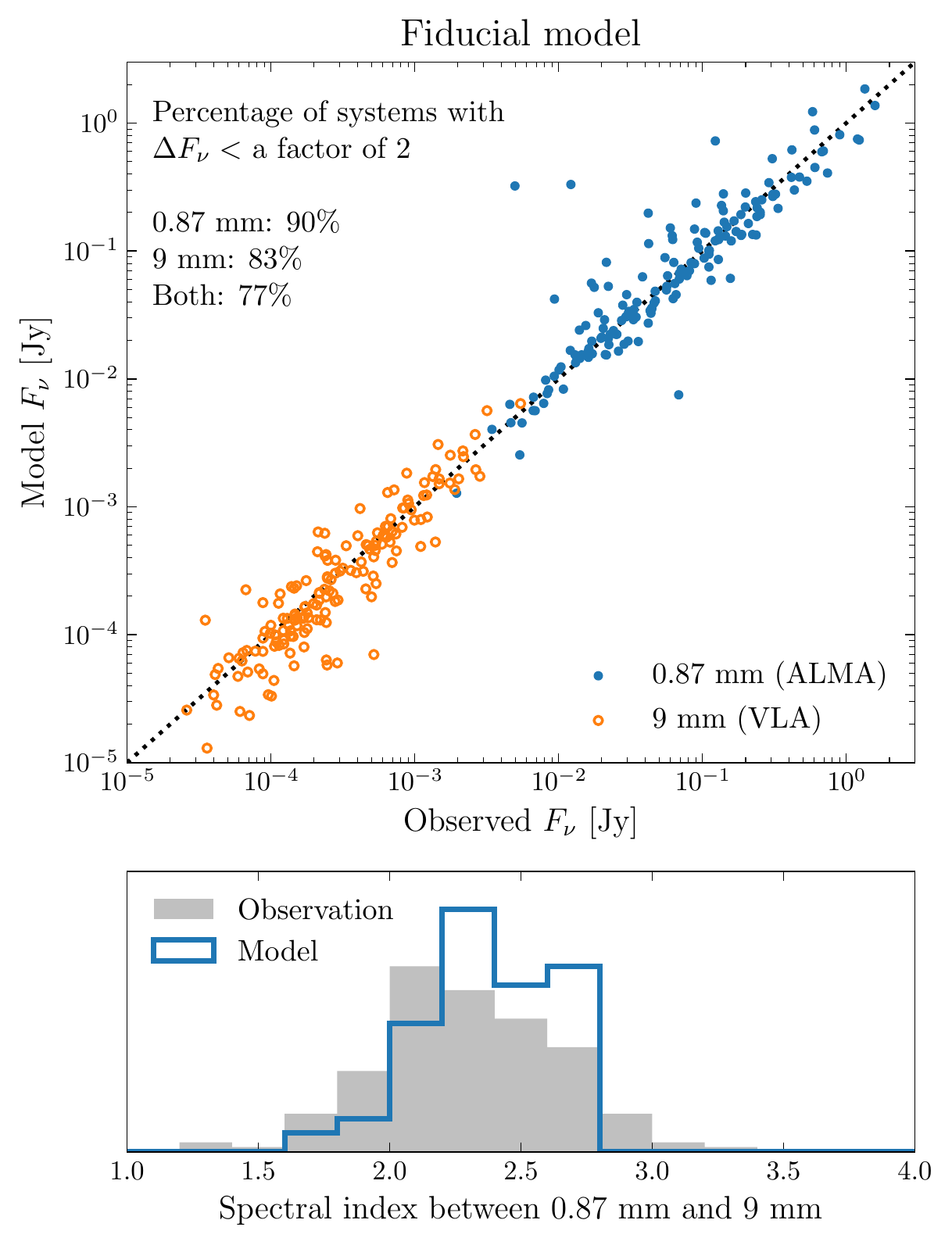}
    \caption{Top panel: Comparison between flux densities from our model and observation. The observed flux densities are quoted from \citetalias{Tobin2020}. Bottom panel: the distribution of spectral indices. For most of the systems, our model is able to fit the flux at both wavelengths well.}
    \label{fig:flux_comp_fid}
\end{figure}

\begin{figure}
    \centering
    \includegraphics[scale=0.66]{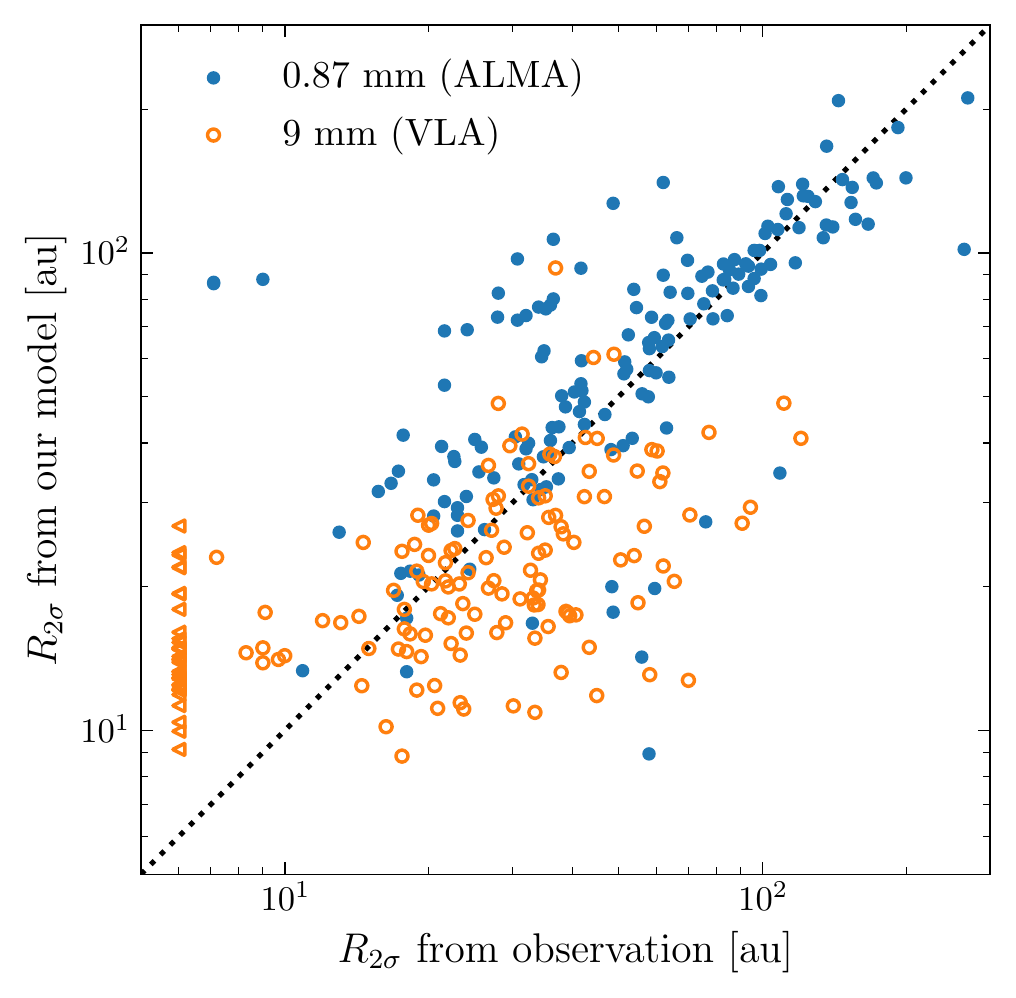}
    \caption{Comparison between apparent disk sizes from our model and observation. Here the apparent disk size $R_{2\sigma}$ is $2\sigma$ of the deconvolved gaussian fit to the observed image or mock observation. Our model broadly agrees with observation at both wavelengths, and reproduces the trend that $R_{2\sigma}$ is systematically smaller at longer wavelength.}
    \label{fig:R2sigma_comp}
\end{figure}

\begin{figure}
    \centering
    \includegraphics[scale=0.66]{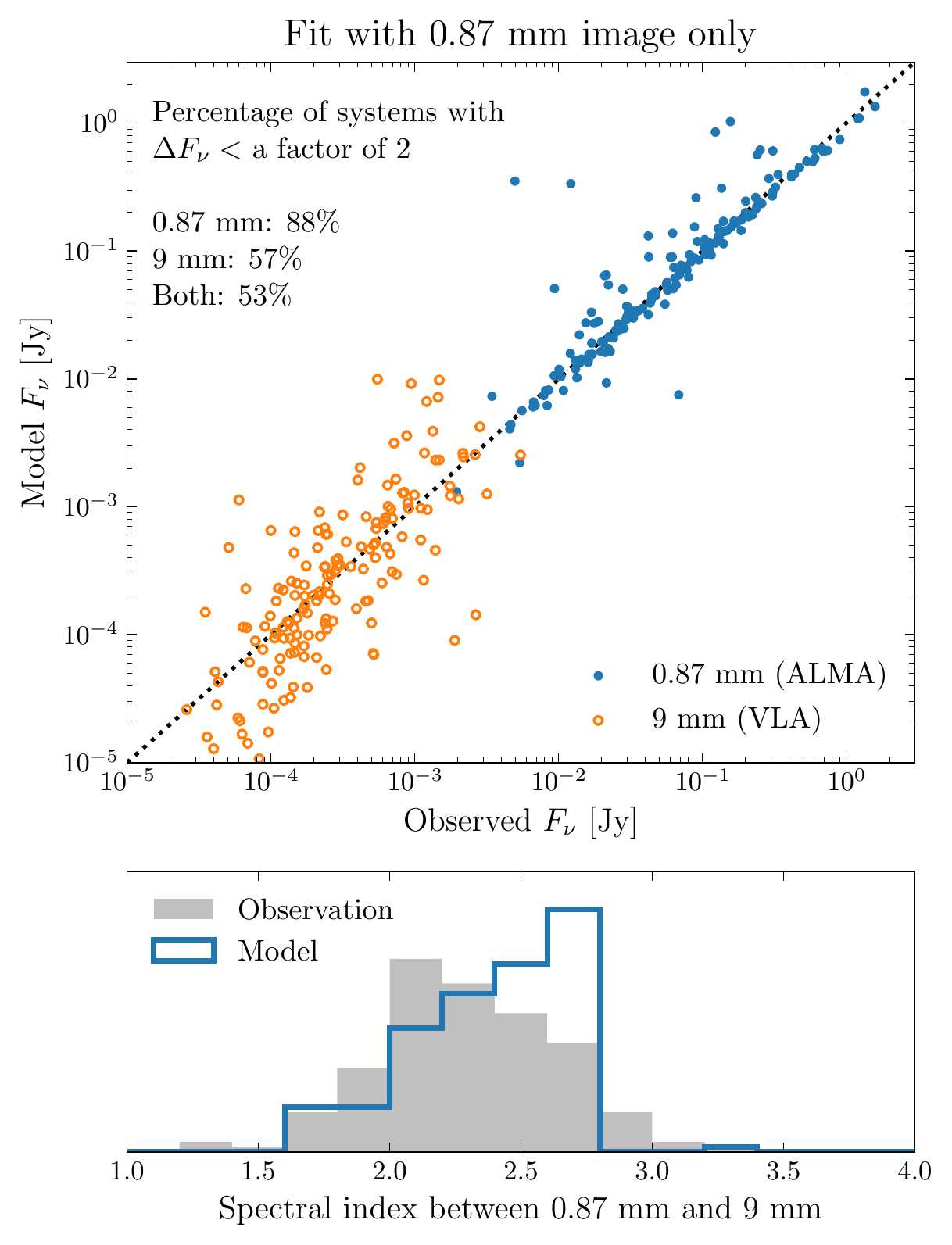}
    \caption{Same as Fig. \ref{fig:flux_comp_fid} but shows results for our model fitted using only the 0.87~mm image. There is more scattering for the 9~mm flux but no systematic difference between model prediction and observation. This demonstrates the predictive power of our model, as most of the 9~mm flux come from dust that is not visible ($\tau\gg 1$) at 0.87~mm.}
    \label{fig:flux_comp_alma}
\end{figure}


\begin{figure}
    \centering
    \includegraphics[scale=0.66]{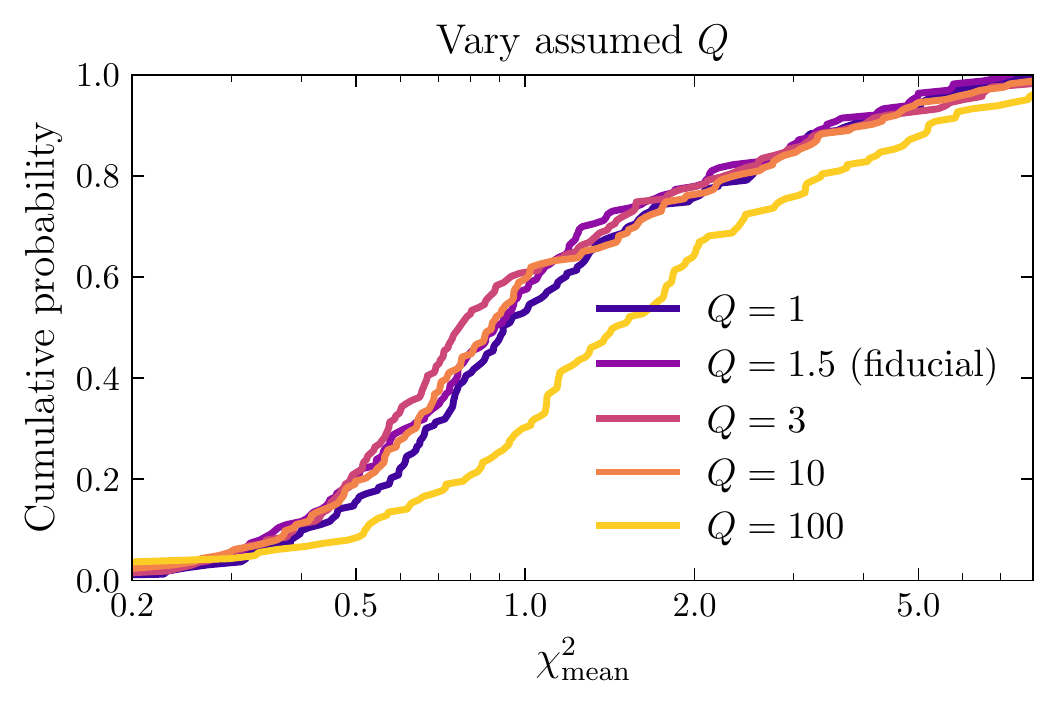}
        \caption{Cumulative distribution of model error $\chi_{\rm mean}^2$ when we vary the assumed Toomre $Q$. Observation strongly prefers $Q\lesssim 10$ over $Q\sim 100$.}
    \label{fig:chi_sq_Q}
\end{figure}

\begin{figure}
    \centering
    \includegraphics[scale=0.66]{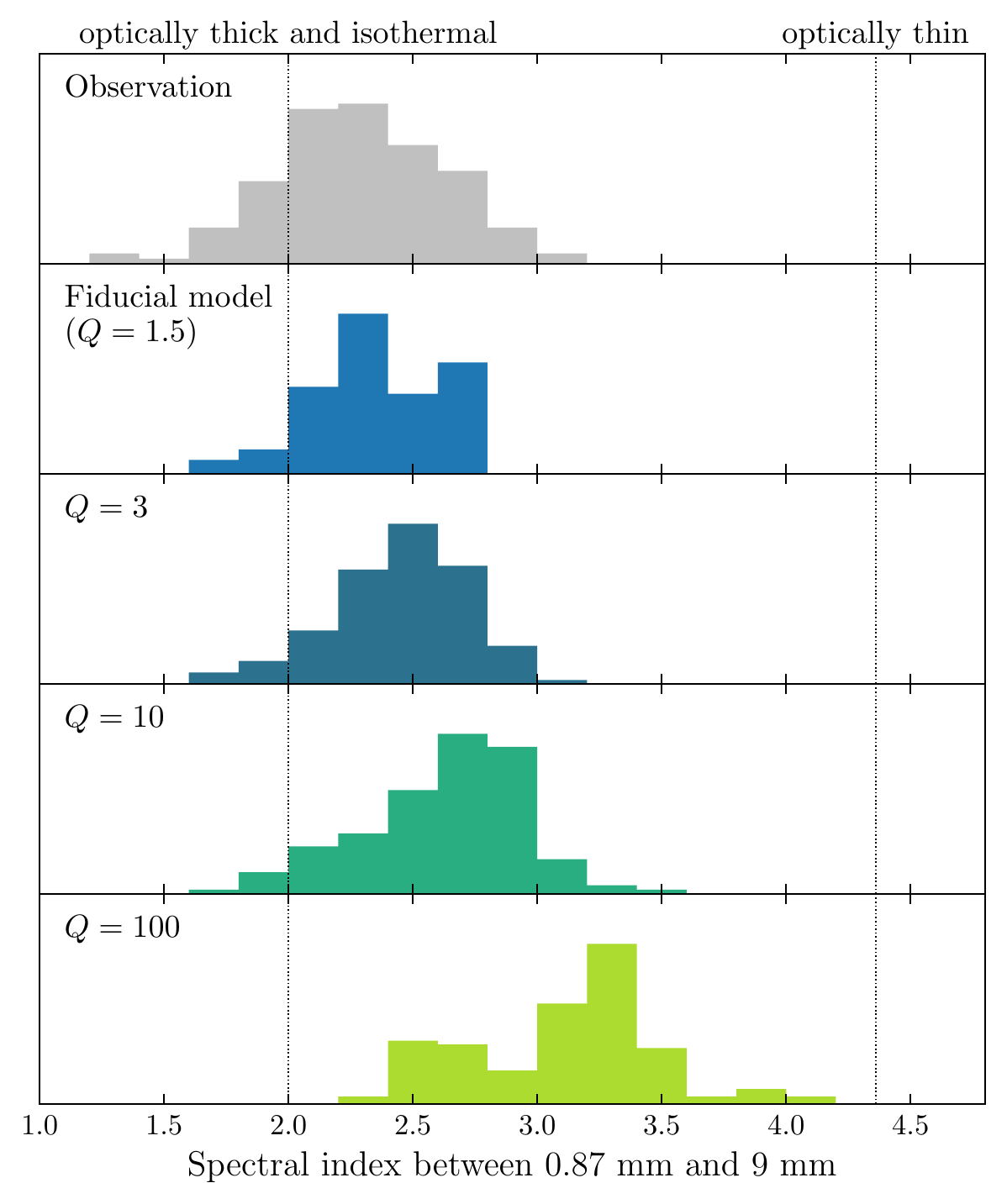}
    \caption{Distribution of disk spectral indices for observed systems (first panel), our fiducial model (second panel), and models assuming larger $Q$ values (last three panels). At large $Q$, the model is no longer able to reproduce the observed spectral index distribution, even though observations at both wavelengths are used to fit the model.}
    \label{fig:spectral_index_Q}
\end{figure}

\subsection{Agreement between model and observation}\label{sec:test_agreement}
In order to test our model against observation, we first check whether the model is consistent with a large fraction of systems in our sample. Note that we do not expect the model to be consistent with all systems, as the physical assumptions of our model corresponds to only one of several possible scenarios of disk formation (see Section \ref{sec:other_formation_scenarios}).

We evaluate the agreement between the best-fit model and observation using the mean $\chi^2$ inside each disk, 
\begin{equation}
\chi_{\rm mean}^2 = \frac{\chi_A^2 S_A S_{B,A}^{-1} + \chi_V^2 S_V S_{B,V}^{-1}}{S_AS_{B,A}^{-1}+S_VS_{B,V}^{-1}}.
\end{equation}
Here $S_{A/V}$ is the area of the region where the ALMA/VLA mock observation is above the detection limit, and $\chi_{A/V}^2$ is the mean $\chi^2$ for the ALMA/VLA mock observation within this region.
In other words, $\chi_{\rm mean}^2$ is an average of $\chi_{A/V}^2$ weighted by the (detectable) disk area in unit of beams.
Fig. \ref{fig:compare_fit_with_img} provides a few examples of systems at different $\chi_{\rm mean}^2$. For $\chi_{\rm mean}^2\lesssim 1$, the model agrees well with observation.
Fig. \ref{fig:chi_sq_fid} plots the distribution of $\chi_{\rm mean}^2$ for our fiducial model; $57\%$ of our sample can be fit reasonably well by the model ($\chi_{\rm mean}^2 \leq 1$).

Fig. \ref{fig:chi_sq_Mdot_and_amax} shows the distribution of $\chi_{\rm mean}^2$ if we vary the assumed dust-size distribution and accretion rate, which are both subject to relatively large systematic uncertainties. The agreement between model and observation is relatively insensitive to different choices of these parameters, while showing a slight preference for a maximum grain size $a_{\rm max} \sim$~mm over larger (1~cm) or smaller (${\lesssim}100~\mu$m) grains.

To give a more direct impression on how well the model fits observation, we also compare flux density (Fig. \ref{fig:flux_comp_fid}) and apparent disk size (Fig. \ref{fig:R2sigma_comp}) between model and observation. Our model is consistent with observations at both wavelengths, and there is no apparent systematic difference in any of these metrics. Especially, our model naturally reproduces the trend that the apparent disk size shrinks towards longer wavelength, which is also visible in Fig. \ref{fig:compare_fit_with_img}.

\subsection{Is the good agreement coincidental?}

Good agreement between observation and model prediction alone cannot be a strong evidence for arguing that the model is a good description of real disks. It remains possible that the good agreement is just an overfit, where the model is too flexible (or the observables provide too few constraints) and predicts an unrealistic disk profile which happens to produce the right observables. While one cannot completely rule out this possibility, we argue that an overfit is unlikely as follows.

First, we note that our model has only two degrees of freedom (two free parameters, $M_\star$ and $R_{\rm d}$), while the observations often offer more constraints than that. For disks that are well resolved by ALMA, the model needs to fit not only the integrated flux at both wavelengths but also the disk size and the radial profile of the flux density. In Fig. \ref{fig:chi_sq_fid} we see that the agreement between the model and well-resolved systems remains good, suggesting that the agreement between the model and observation is likely not because the model has too many free parameters.

Second, we test the predictability of our model by fitting it only with single-wavelength 0.87-mm (ALMA) observations and use it to predict the 9-mm flux. This is not a trivial task; the emission is generally optically thick at 0.87~mm, and most of the 9-mm emission come from dust that is not visible at 0.87~mm. Therefore, correctly predicting the 9-mm flux requires a good estimate of the total dust mass using the emission at the surface of the disk and the assumed physical constraints in our model. In Fig. \ref{fig:flux_comp_alma} we see that the model predicts the 9-mm flux without systematic error and reproduces the typical spectral index relatively well.

\subsection{Does observation favor gravitationally self-regulated disks?}

A central assumption of our model is that the disk is gravitationally self-regulated with a marginally unstable Toomre $Q$.
However, as we commented earlier, this assumption should not be taken for granted (cf. Section \ref{sec:other_formation_scenarios}).
Here we try to use observation to constrain (at the population level) the typical Toomre $Q$ parameter of our sample.

In Fig. \ref{fig:chi_sq_Q} we compare the distribution of $\chi_{\rm mean}^2$ for several different assumed values of $Q$. The agreement between observation and model is similarly good for $Q=1-10$, but deteriorates as we further increase $Q$.
Fig. \ref{fig:spectral_index_Q} compares the distribution of disk spectral index for different $Q$ values; it shows a preference for $Q=1.5$ over larger $Q$ values, with the caveat that the difference between $Q=1.5$ and $Q=3$ remains small.

In summary, these evidences show that order-unity $Q$ is preferred at a population level. We can understand this preference by noticing that, at a higher $Q$ and for the same (optically thick) 0.87-mm emission, the model would under-predict the surface density and 9-mm flux, making it difficult to reproduce the observation at both wavelengths.

\section{Disk properties and physical implications}\label{sec:results}

\begin{figure}
    \centering
    \includegraphics[scale=0.66]{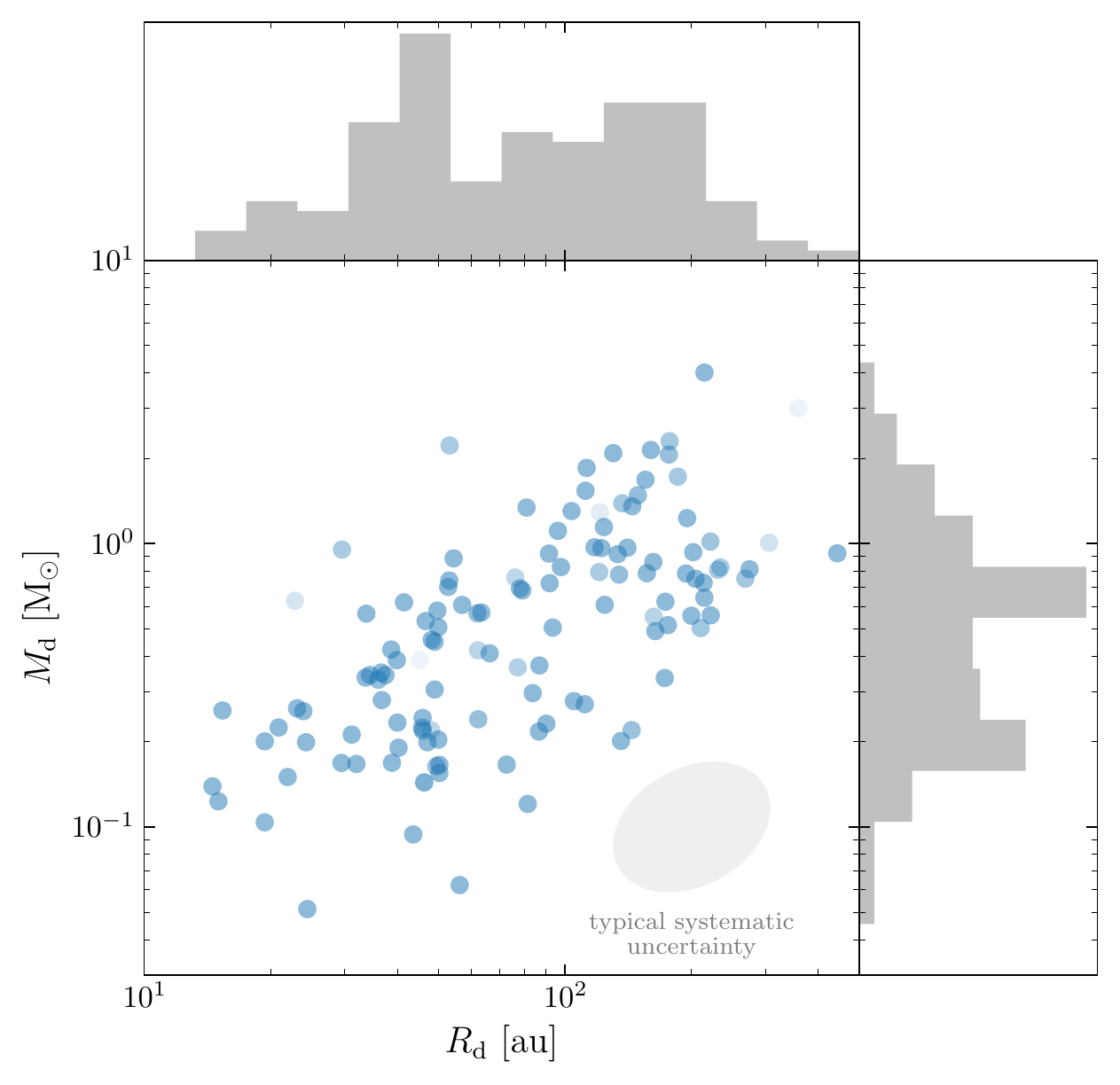}
    \caption{Distribution of disk size $R_{\rm d}$ and mass $M_{\rm d}$ for our fiducial model. We only include systems with $\chi_{\rm mean}^2\leq 2$; for $\chi_{\rm mean}^2>1$, the opacity of the marker decreases as $\chi_{\rm mean}^2$ increases. We also plot the $1\sigma$ uncertainty due to systematic uncertainties in the assumed model parameters for reference (cf. Appendix \ref{a:uncertainty}). There is a positive correlation between disk size and mass.}
    \label{fig:scatter_Rd_Md}
\end{figure}

\begin{figure}
    \centering
    \includegraphics[scale=0.66]{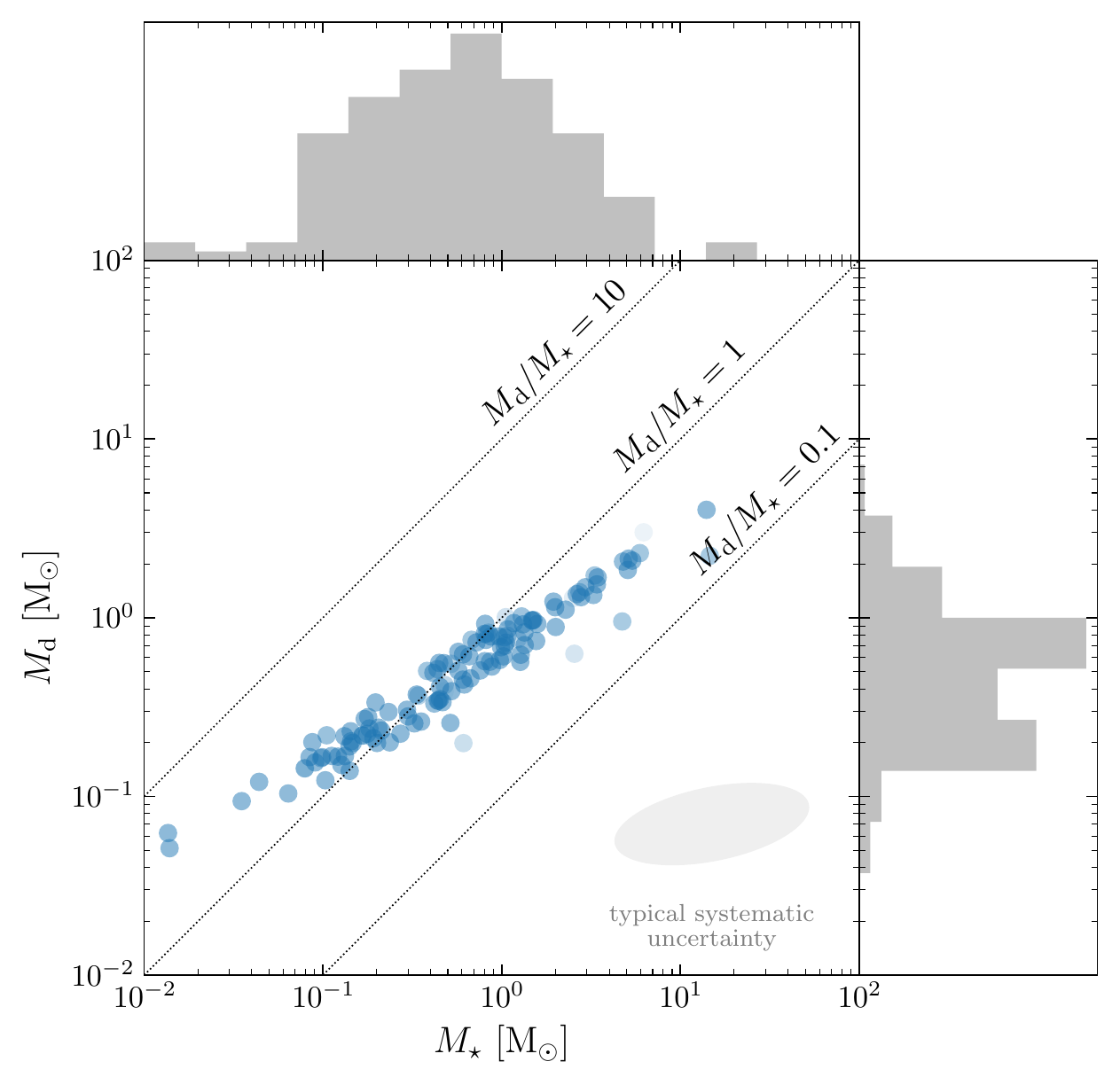}
    \caption{Same as Fig. \ref{fig:scatter_Rd_Md}, but for the distribution of stellar mass $M_\star$ and disk mass $M_{\rm d}$. Most systems have order-unity disk-to-star mass ratio. There appears to be a strong correlation between $M_{\rm d}$ and $M_\star$, with slope $\approx 0.6$; this is a generic property of gravitationally self-regulated disks (Section \ref{sec:results_summary}; cf. Appendix \ref{a:mass_scaling}).}
    \label{fig:scatter_Ms_Md}
\end{figure}

\begin{deluxetable*}{l @{\hspace{2em}} cccc @{\hspace{2em}} ccc @{\hspace{2em}} ccc}
\tablecaption{Distributions of key disk properties}
\startdata\\
~ & \multicolumn{4}{c@{\hspace{2em}}}{$M_{\rm d}$  [${\rm M}_\odot$]} & \multicolumn{3}{c@{\hspace{2em}}}{$R_{\rm d}$  [au]} & \multicolumn{3}{c@{\hspace{2em}}}{$T_{\rm mean}$  [K]} \\
~ & Mean & 25\% & Med & 75\% & 25\% & Med & 75\% & 25\% & Med & 75\% \\
\tableline
All  & 0.68 & 0.23 & 0.52 & 0.85 & 45.3 & 77.7 & 144.4 & 228.0 & 272.4 & 311.6 \\
Class 0  & 0.72 & 0.26 & 0.62 & 0.93 & 48.8 & 92.8 & 174.1 & 221.4 & 268.7 & 299.8 \\
Class I  & 0.73 & 0.30 & 0.45 & 0.87 & 36.7 & 54.4 & 119.1 & 268.1 & 305.5 & 331.4 \\
Flat Spectrum  & 0.39 & 0.17 & 0.24 & 0.40 & 44.0 & 54.6 & 78.9 & 238.6 & 253.3 & 265.7 \\
\tableline
T20 (0.87 mm)  & 0.024 & 0.005 & 0.013 & 0.031 & 32.5 & 51.8 & 82.9 & 44.0 & 58.0 & 71.9 \\
\enddata
\tablecomments{Here we only include systems with $\chi_{\rm mean}^2\leq 2$.
The last row shows results from \citetalias{Tobin2020} (based on 0.87 mm observation) for the same sample.}
\label{tab:disk_properties}
\end{deluxetable*}

\begin{figure}
    \centering
    \includegraphics[scale=0.66]{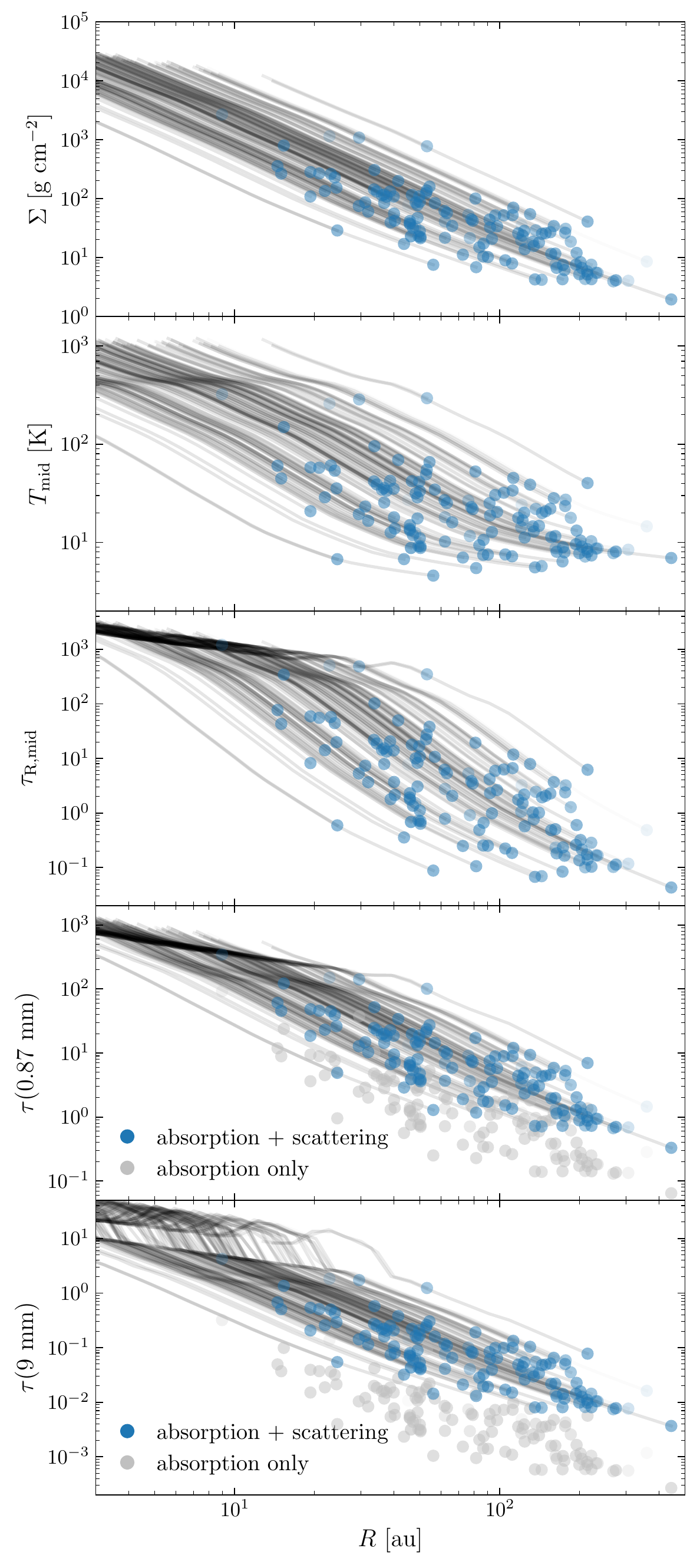}
    \caption{Radial profiles of disks in our sample, estimated with our fiducial model. From top to bottom, we show surface density $\Sigma$, midplane temperature $T_{\rm mid}$, miplane Rosseland mean optical depth $\tau_{\rm R, mid}$, and optical depth at the two observed wavelengths. We mark the disk edge with a blue dot in each curve. For the last two panels we also show the optical depth at disk edge for absorption opacity only. We only include systems with $\chi_{\rm mean}^2\leq 2$; for $\chi_{\rm mean}^2>1$, the opacity of the marker decreases as $\chi_{\rm mean}^2$ increases.}
    \label{fig:radial_profile}
\end{figure}


In this section we discuss the new estimates of the properties of Orion protostellar disks obtained with our fiducial model and discuss the physical implication of our results on disk evolution and fragmentation.

\subsection{Summary of disk properties and comparison with previous studies}\label{sec:results_summary}

We begin with an overview of the estimated properties of our disks, which are summarized in Table \ref{tab:disk_properties} and Figs. \ref{fig:scatter_Rd_Md} -  \ref{fig:radial_profile}.
Here we only focus on systems with $\chi_{\rm mean}^2\leq 2$, as for systems with large $\chi_{\rm mean}^2$ our model may not be able to provide reliable estimates.

\textbf{Disk mass and surface density:} Our sample has a mean disk mass of 0.68 ${\rm M}_\odot$ and median disk mass 0.52 ${\rm M}_\odot$, which is often comparable to or more massive than the protostar mass; there also appears to be a correlation between disk mass and stellar mass, with slope $\approx 0.6$ (Fig. \ref{fig:scatter_Ms_Md}). The surface density profile of our disks approximately scale as $\Sigma\propto r^{-2}$ (Fig. \ref{fig:radial_profile}), therefore a significant portion of disk mass is concentrated in the inner part of the disk. These properties are all generic features of a gravitationally self-regulated disk, and agrees with previous semi-analytic calculations (e.g., \citealt{LinPringle1990, Rafikov2009} also see Appendix \ref{a:mass_scaling}). Also note that the tight correlation in Fig. \ref{fig:scatter_Ms_Md} partly because we have assumed (for simplicity) consant $\dot M\propto M_\star$. In reality, the distribution would likely be more scattered.

Our mass estimates are significantly higher than previous estimates for the same data set (\citetalias{Tobin2020}, \citealt{Sheehan2022}). Such difference is mainly due to different model assumptions; we make a more detailed comparison between these models and argue that our model features better predictive power in Section \ref{sec:model_comparison}.

Our disks are also somewhat more massive than the disks in the large-scale hydrodynamics disk population synthesis by \citet{Bate2018}, which gives $M_{\rm d}/M_\star\sim 0.1-1$. This difference could be related to insufficient resolution as commented in the resolution study in \citet{Bate2018}. \citetalias{XK21b} also demonstrated that $M_{\rm d}/M_\star$ can be $\mathcal O(1)$ even when magnetic fields have been included, as long as there is no excessive numerical dissipation (which might be common among earlier simulations due to their relatively low resolution in the innermost several 10~au).

\textbf{Disk size:} The disk size in our sample has a median of 77.7~au and shows a wide distribution, with a factor of ${\sim}3$ difference between the first and third quartile. Our disk sizes are often larger than those quoted in \citetalias{Tobin2020}, but this is mainly because \citetalias{Tobin2020} is reporting the apparent disk size $R_{2\sigma}$; there is no longer a systematic difference if we compare $R_{2\sigma}$ of our model with the \citetalias{Tobin2020} results (Fig. \ref{fig:R2sigma_comp}). We also see a positive correlation between disk mass and disk size; similar correlation has been reported for older disks \citep{Tripathi2017}.

\textbf{Disk temperature:} The midplane temperature in our disks range from ${\lesssim}10$~K to 1200~K, and the mean temperature $T_{\rm mean}$ are generally several 100~K. Such high temperature is in part because the disk is often optically thick ($\tau_{\rm R,mid}\gtrsim 1$) and cannot cool efficiently. It is also related to the steep surface density scaling, which makes the hotter inner disk dominate the density-weighted averaging. The disk temperature decreases in radius with a relatively steep power-law slope ${\approx}-1$, and $T_{\rm mid}$ generally varies by more than an order of magnitude across the disk. (As a result, $T_{\rm mean}$ should not be interpreted as a single characteristic temperature of the disk.) Also note that the vertical temperature profile (Eq. \ref{eq:T_profile}) implies that the observed dust (from disk surface) can be significantly cooler than $T_{\rm mid}$, especially at 0.87~mm.

\subsection{Implication on disk evolution}\label{sec:evolution_trends}

\begin{figure}
    \centering
    \includegraphics[scale=0.66]{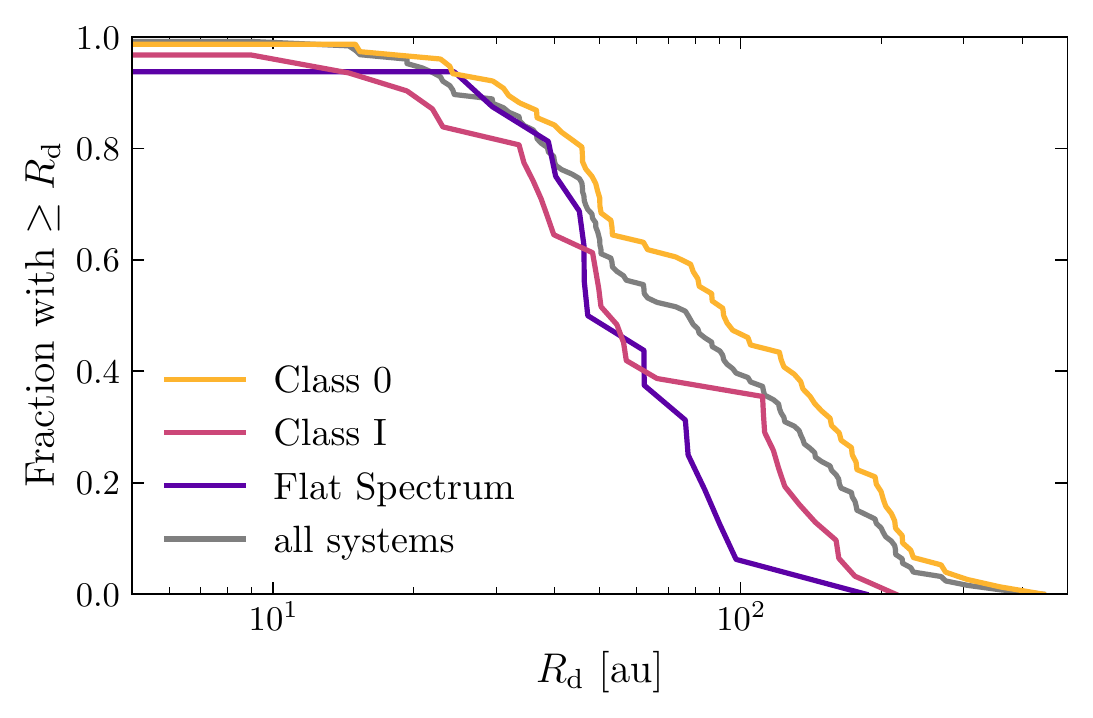}
    \includegraphics[scale=0.66]{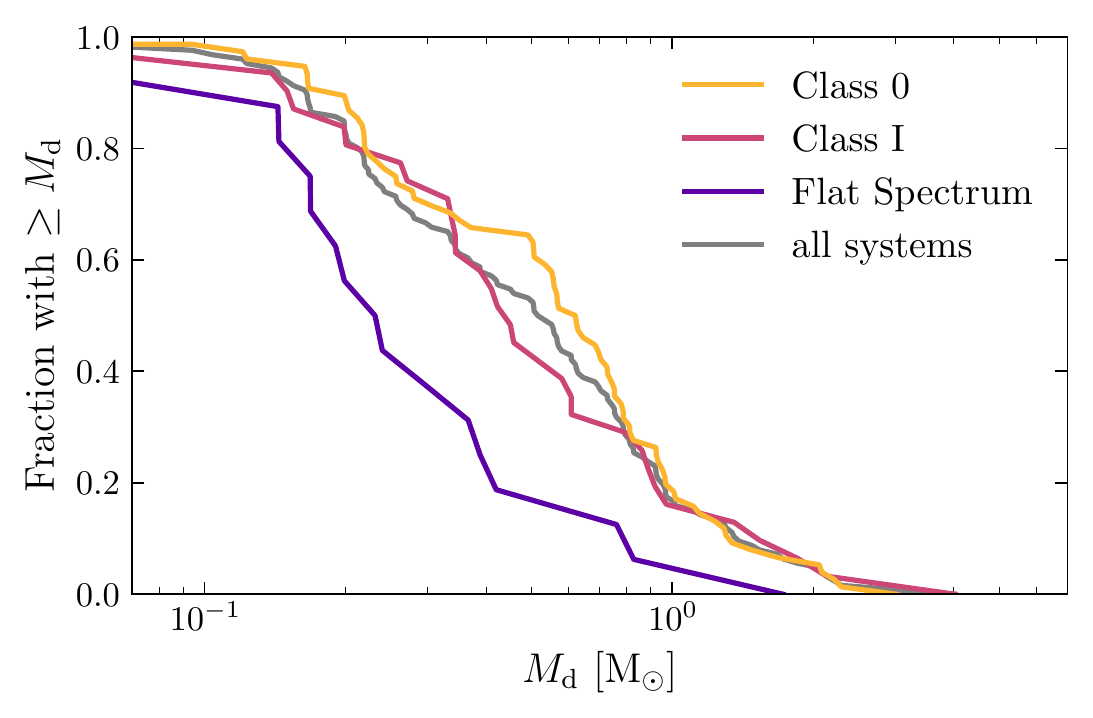}
    \caption{Cumulative distribution of disk size (top panel) and mass (bottom panel) for different evolutionary stages. Both disk size and mass decrease towards later evolutionary stages. These trends are discussed in Section \ref{sec:evolution_trends}.}
    \label{fig:class}
\end{figure}

Now we discuss the evolution of disk size and mass by comparing their statistics across different evolutionary stages and against Class II disks in the literature. We also attempt to link the observed trends with theories of Class 0/I disk evolution.

\textbf{Disk mass evolution:} The estimated masses of Class 0 and Class I disks in our sample are largely similar, while those of Flat Spectrum disks are lower by more than a factor of 2.\footnote{This relation could be biased by our assumption of a constant $\dot M/M_\star$, as in reality $\dot M/M_\star$ should systematically decrease toward later stages. But given the weak correlation between assumed $\dot M/M_\star$ and estimated $M_{\rm d}$ (Table \ref{tab:sensitivity}), this qualitaive trend of $M_{\rm d}$ evolution should remain the same.} Physically, the decrease of disk mass towards later evolutionary stages is expected because as the envelope disperses, the accretion rate could drop exponentially \citep{Fischer2017} and that leads to less accretion heating, lower disk temperature, and a lower disk mass required for gravitational self-regulation. In other words, as the accretion rate drops, a gravitationally self-regulated disk tends to decrease its mass to maintain marginal instability (and this is achieved by keeping the accretion rate from the disk to the star slightly above the accretion rate from the envelope to the disk).

One could also estimate the mass of a gravitationally self-regulated disk at the end of envelope dispersal as an initial condition of Class II evolution. At that point accretion heating is low enough that the disk temperature should become comparable to the ambient temperature of $\sim 10$~K, and that produces a disk mass of order $0.1{\rm M}_\odot$ for a solar-mass star with a $\sim 100$~au disk (cf. \citealt{XK21a} Section 5.3).
This is still significantly higher than early observational estimates of Class II disk masses in young ($\lesssim 3$ Myr) star-forming regions (Taurus, Ophiuchus, Lupus, ONC), which are of order $3\times 10^{-3}{\rm M}_\odot$ (\citealt{Andrews2013,Ansdell2016,Tripathi2017,Eisner2018}; see a summary in \citetalias{Tobin2020} Figs. 14, 15). But the simple models applied in these studies could systematically underestimate the disk mass by 1-2 orders of magnitude according to recent studies adopting different (and probably more robust) methods of mass estimation \citep{Booth2019,Powell2019,Anderson2022}. Accounting for this bias, the typical disk mass at the beginning of Class II would be broadly consistent with gravitationally self-regulated evolution during Class 0/I.

\textbf{Disk size evolution:} The size of disks in our sample decreases after Class 0, consistent with the trend observed in \citetalias{Tobin2020}. While there is no firm conclusion on what causes this trend, one possible explanation for the shrinking disk size is enhanced magnetic braking in the inner envelope at later times \citepalias{XK21b}. During the collapse of a magnetized core, ambipolar diffusion decouples magnetic flux from the gas before the gas is accreted by the protostar-disk system. Most of the magnetic flux that initially belongs to the material in the protostellar-disk system piles up around it in a growing magnetically dominated region (a smoother version of the ``magnetic wall'', cf. \citealt{LiMcKee1996,TassisMouschovias2005}).
At later times, infalling material needs to pass through this magnetically dominated region and would lose most of its angular momentum before reaching the disk. In this case, while accretion is still increasing the total mass of the protostar-disk system ($M_\star+M_{\rm d}$), it barely increases the angular-momentum budget of the disk. This could lead to a decrease in disk size, as demonstrated in the simulation of \citetalias{XK21b}.

This trend of shrinking disk size should end as the envelope disperses and system transitions into Class II, and the disk size can start growing again if the disk remains mainly regulated by angular-momentum transport (e.g., gravitational or magneto-rotational instability) as opposed to angular-momentum removal (wind, magnetic braking).\footnote{Here ``disk size'' refers to the gas disk size. The dust disk size generally decreases, because unlike Class 0/I disks (Appendix \ref{a:grain_coupling}), Class II disks show a significant level of radial drift of dust grains \citep[cf.][]{Ansdell2018} due to their longer evolution timescale and weaker dust-gas coupling.}
Unfortunately, it would be difficult to test whether disk size increases during the transition into Class II with observation. While Class II disk-size estimates are available for a number of young (${\lesssim} 3$ Myr) star-forming regions \citep{Eisner2018}, the typical disk size can differ by a factor of a few across different star-forming regions, making it difficult to make a meaningful comparison between these populations and our sample of Orion protostellar disks.

\textbf{Summary:} Comparing disk properties across different stages of protostellar evolution and with statistics of Class II disks in the literature, we find that the evolution is affected by several different (and often competing) mechanisms, resulting in non-monotonic evolution where the evolution of disk size and mass switches from growth to decay before finishing Class 0. Especially, the cooling of the disk during the dispersal of the envelope might play an important role in bridging protostellar disks to young protoplanetary disks, which are significantly less massive.

\subsection{Fragmentation}\label{sec:fragmentation}

\begin{figure}
    \centering
    \includegraphics[scale=0.66]{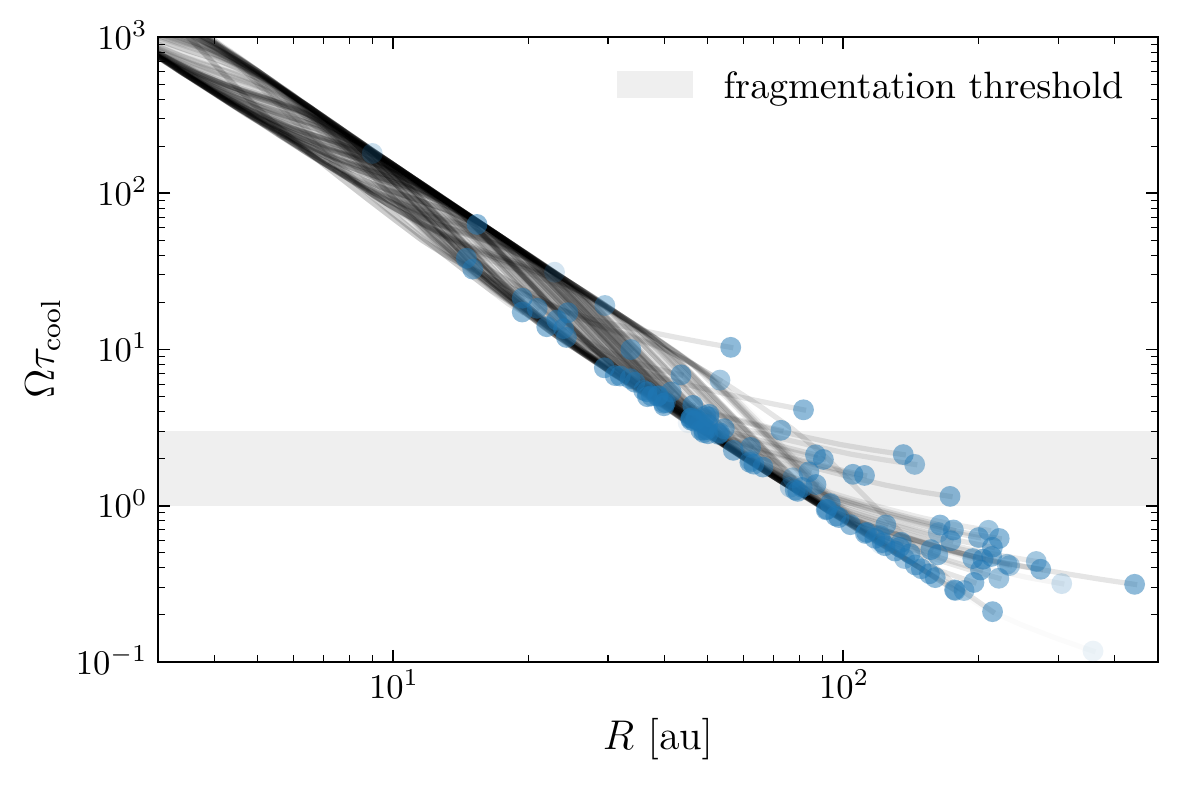}
    \caption{$\Omega\tau_{\rm cool}$ profiles of our best-fit models. 40-60\% of the disks in our sample are likely prone to fragmentation beyond 50-100~au, with the caveat that our model may not be applicable to fragmenting disks. See discussion in Section \ref{sec:fragmentation}.}
    \label{fig:Omega_tau_cool}
\end{figure}

One important possible outcome of GI is fragmentation, which can lead to the formation of (stellar and sub-stellar) companions and are thought to be related to the outbursts which are common among protostellar systems \citep{VB06}. Here we check whether the disks in our sample might be prone to fragmentation.

Fragmentation occurs when the overdensities produced in GI-driven perturbations cool rapidly enough to collapse under their own gravity before being disrupted by orbital shear and other perturbations. Traditionally, the condition of fragmentation is $\Omega\tau_{\rm cool} < 1-3$ \citep{Gammie2001}. Here the cooling timescale $\tau_{\rm cool}$ is the ratio between disk internal energy $U$ and cooling rate $2\sigma T_{\rm eff}^4$, both of which can be computed from our model disk profile.

In Fig. \ref{fig:Omega_tau_cool} we plot the radial profile of $\Omega\tau_{\rm cool}$ for our best-fit models.
$\Omega\tau_{\rm cool}$ decreases in radius, and 40-60\% of the disks in our sample are prone to fragmentation beyond 50-100 au.
This is broadly consistent with the analytic prediction of \citet{Clarke2009} and the observation of a bimodal distribution of companion separation in protostellar systems \citep{Tobin2016,Tobin2022} where the peak at smaller separation (${\sim}100$~au) could be due to fragmentation.
Note that such observation does not provide a very good constraint on whether fragmentation commonly occurs as the visibility and survival rate of fragments are still not well understood.

It is worth noting that our physical picture of a gravitationally self-regulated disk cannot be directly applied to a fragmenting disk.
Still, it is possible that a similar kind of self-regulation exists whereby the disk self-regulates by switching between gravitationally stable and unstable (fragmenting) states \citep[cf.][]{VB13, Vorobyov2019}. In this case, whenever the surface density becomes high enough for disk to be unstable and fragment, this part of the disk can get rid of some mass by forming this fragment which will migrate away, and that (together with the associated angular-momentum transport and heating) can push the disk back towards a gravitationally stable state. This fragmentation self-regulation keeps the disk around a marginally stable state, with occasional fragmentation. Qualitatively, this is similar to the constraint of gravitational self-regulation used in our model. One caveat is that this argument has not been studied in sufficient detail in theory or simulation to allow reliable predictions, and there is still a lot of uncertainties regarding the interplay between fragmentation and disk evolution.

\section{Comparison with previous studies}\label{sec:model_comparison}

\begin{figure}
    \centering
    \includegraphics[scale=0.66]{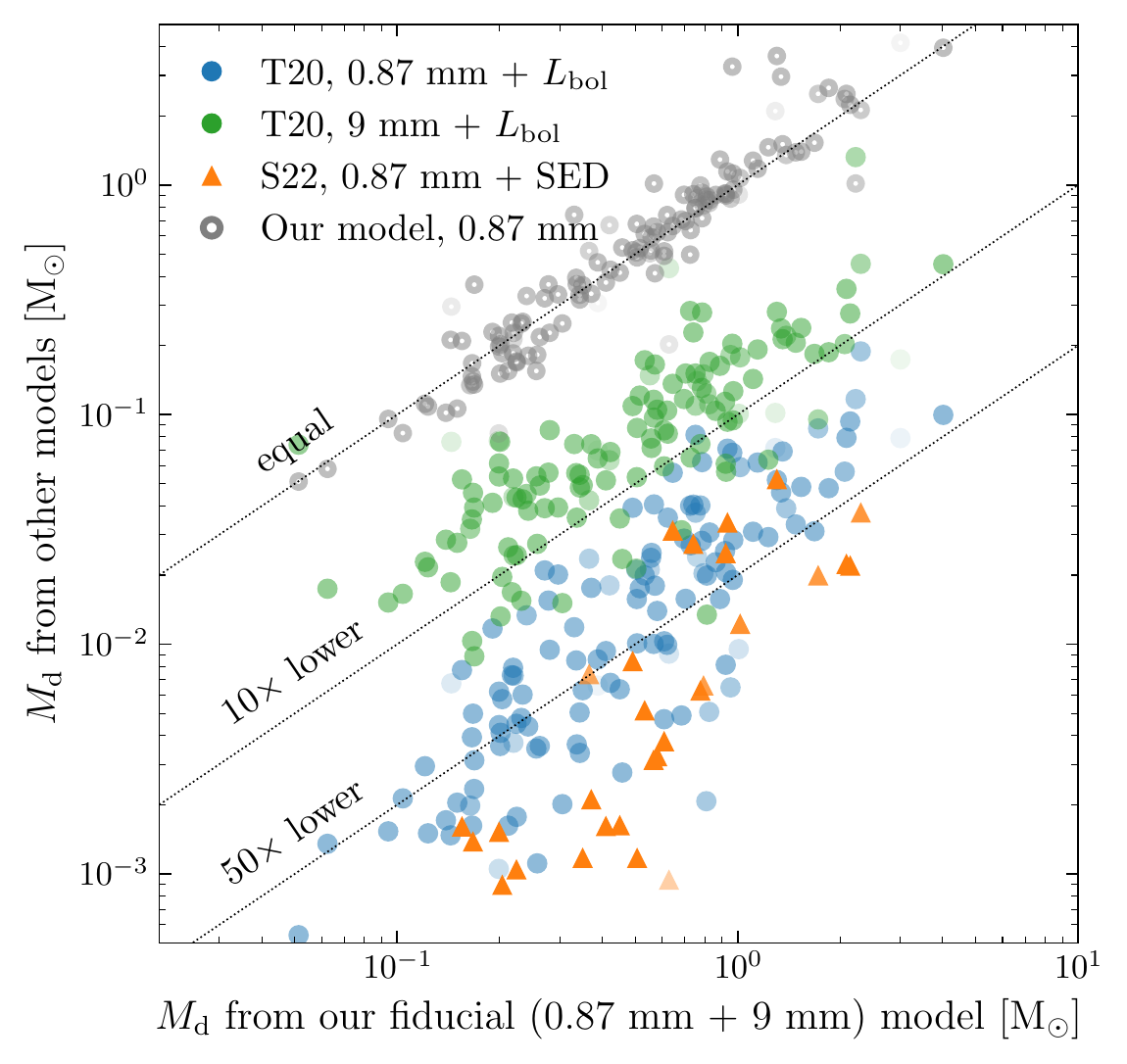}
    \caption{Comparison between disk mass estimates from our fiducial model (fitted with images at both wavelengths) and models based on images at a single wavelength. These include the optically thin, isothermal models from \citetalias{Tobin2020} for each observed wavelength (blue and green dots), the parametrized disk+envelope model coupled with radiative transfer from \citet{Sheehan2022} (orange triangles), and our model fitted with 0.87~mm images only (grey circles). The difference can be attributed to the fact that single-wavelength observation leaves a degeneracy between mass (or optical depth) and temperature, which need to be resolved by the (highly different) physical assumptions of each model (Section \ref{sec:model_comparison}).}
    \label{fig:scatter_vandam_M}
\end{figure}

Our disk model produces disk masses that are significantly higher than those in \citetalias{Tobin2020} and \citet{Sheehan2022}, as shown in Fig. \ref{fig:scatter_vandam_M}. In this section we discuss why it is possible to get such drastically different estimates from the same observational constraints, and point out that the prediction of our model is more consistent with the constraints from multi-wavelength observation.

\subsection{Summary of models}

We begin by summarizing the model used in these three studies. \citetalias{Tobin2020} estimated the disk mass by assuming isothermal, optically thin dust; hence the disk mass is directly porportional to the observed flux. The dust temperature was assigned following an empirical scaling law, $T_{\rm dust} = 43\,({\rm L}_{\rm bol}/L_\odot)^{1/4}$~K. Since this estimate requires only single-wavelength flux, \citetalias{Tobin2020} obtained separate disk mass estimates for 0.87 mm and 9 mm; the two estimates differ by a factor of $\sim 5$.\footnote{\citetalias{Tobin2020} also adopted a much higher 9-mm opacity, which comes from assuming a mm-cm dust opacity index of $\approx 1$; such small dust opacity index requires the grains to be highly porous \citep{Woitke2016} or have size much larger than this wavelength (i.e. $\gtrsim$ a few cm; cf. \citealt{Draine2006}). The difference between \citetalias{Tobin2020}'s estimates at 0.87 mm and 9 mm would be more significant (and their 9 mm estimate would be closer to our mass estimate) if \citetalias{Tobin2020} adopted our dust model.}

\citet{Sheehan2022} performed radiative transfer calculations on a generic, 17-parameter disk$+$envelope model to fit the SEDs (up to 0.87 mm) and 0.87 mm visibilities for a subset of the VANDAM Orion sample. The 9-mm data has not been used for fitting or validation. The model assumes that the grains are mainly heated externally and ignores internal accretion heating. The resulting disks are optically thin at 0.87~mm with masses slightly lower than the \citetalias{Tobin2020} results.

Our model fits the 0.87-mm and 9-mm images simultaneously and assumes that the disk is marginally gravitationally unstable and internally heated; external heating by protostellar irradiation is ignored (Section \ref{sec:physical_assumptions}). Our model has only two free parameters to ensure that the data have more degrees of freedom than the model, and fiducial values are assigned to other relevant parameters (Section \ref{sec:parameter}). We generally find the disks to be massive and optically thick. We also tried to fit our model with only 0.87-mm images and obtained similar results (Fig \ref{fig:flux_comp_alma}; also see grey circles in Fig \ref{fig:scatter_vandam_M}).

\subsection{Why can models get different results from the same observation?}

Fundamentally, this is because the models in \citetalias{Tobin2020} and \citet{Sheehan2022} are both based on single-wavelength images, yet the disk properties are under-constrained by single-wavelength observation. Roughly speaking, the luminosity of the disk is given by
\begin{equation}
L_\nu \sim \kappa_{\nu,\rm abs} M_\nu^{\rm \tau<1} B(T_\nu^{\tau<1})
\end{equation}
where $M_\nu^{\rm \tau<1}$ is the mass of the visible ($\tau_\nu<1$) portion of the disk (or $\tau_\nu\sqrt{1-\omega_\nu}$ if scattering is strong; cf. \citealt{Zhu2019}) and $T_\nu^{\tau<1}$ the typical temperature in this region. If we only know $L_\nu$ (from $F_\nu$), there is still a degeneracy between mass and temperature.

Estimating disk properties requires lifting this degeneracy (or constraining how much mass is invisible, when the disk is optically thick). Each model achieves this by using its own physical assumptions (which could also involve additional observational information such as $L_{\rm bol}$ and SED) to provide additional constraints. In other words, the estimated disk properties would depend on both observational constraints and model assumptions. Therefore, it is not too surprising that models using significantly different physical assumptions produce different estimates of disk properties but all fit the observed data well.




\subsection{Which model should I trust?}

From a theoretical perspective, each of the three models have their limitations. For example, in terms of the dust heating mechanism, \citetalias{Tobin2020} and \citet{Sheehan2022} only consider protostellar heating while we only consider accretion heating; in reality both mechanisms might be important. As a result, it would be difficult to confidently tell \textit{a priori} which of the models is a better approximate to reality. Meanwhile, the three models are all consistent with the observations they are based on; so we would need more stringent tests.

One ideal choice would be to test the model's predictive power, i.e. whether it can predict observables that may not be directly derived from the data fed to the model. This could help to rule out incorrect or overfitted models. The multi-wavelength observations of the VANDAM Orion survey offers an excellent opportunity to perform such test; we can check whether the model can use the 0.87-mm data to predict the 9-mm data (or the spectral index between 0.87 mm and 9 mm).
The models in \citetalias{Tobin2020} and \citep{Sheehan2022} both fail at this test; the produce optically thin disks with dust spectral index $\gtrsim 1$, which results in a spectral index $\gtrsim 3$ (much higher than the observed mean spectral indices of $\sim$2.2) and systematic under-prediction of the 9-mm flux.
Meanwhile, our model demonstrates good predictive power and successfully predicts the 9-mm emission when fitted with only the 0.87-mm data (Fig. \ref{fig:flux_comp_alma}).

In summary, while it is difficult to judge models based on how reasonable their physical assumptions are, current constraints from multi-wavelength observations prefer our model over existing models in \citetalias{Tobin2020} and \citet{Sheehan2022}.

\section{Discussion}\label{sec:discussion}

\subsection{Visibility of GI-induced disk substructures}

\begin{figure}
    \centering
    \includegraphics[scale=.6]{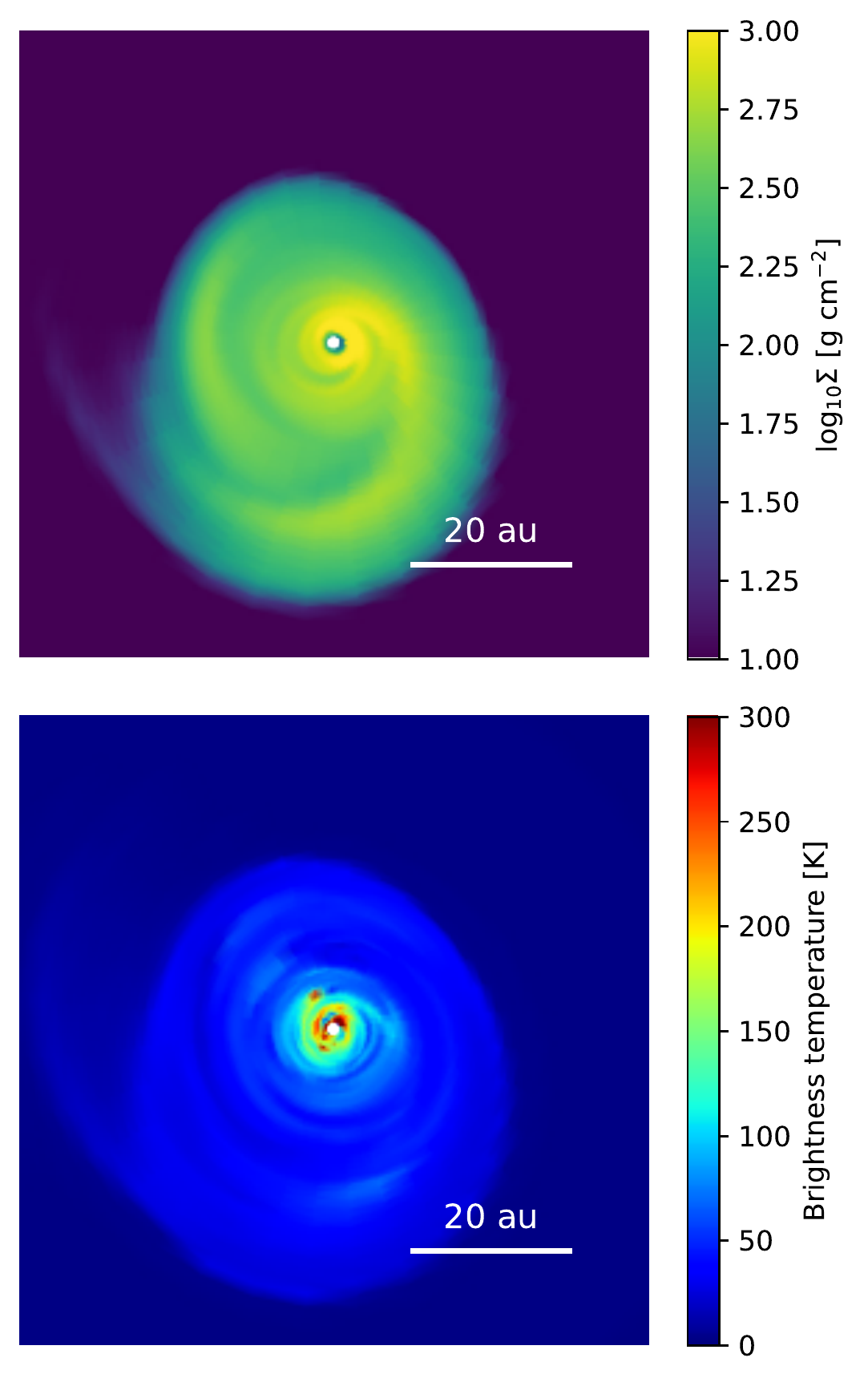}
    \caption{Column density (top panel) and brightness temperature at 0.87~mm (bottom panel; this is proportional to the flux density at this wavelength) from the snapshot of a simulation of protostellar disk formation. While the disk contains prominent large-scale spirals, they are barely visible at sub-mm wavelength due to high optical depth.}
    \label{fig:mock_obs}
\end{figure}

In this subsection we discuss why our argument that the majority of protostellar disks may be gravitationally unstable is not invalidated by the fact that dust-continuum observations of most protostellar disks show no spiral structure. The most straightforward explanation is the lack of resolution, as the typical width of the spirals would be $\sim H$ and our current observation is nowhere close to resolving that. In addition to resolution, the detectability of spirals are also going to be limited by disk optical depth and/or detection sensitivity; here we discuss these issues in order to better plan and interpret future observations.

At sub-mm wavelengths, the high optical depth of the disk would decrease the visibility of gravitationally excited spiral structures.
To provide an example, we re-ran the protostellar disk formation simulation in \citetalias{XK21b} and computed the brightness temperature (which would be directly proportional to flux density) at 0.87~mm in Fig. \ref{fig:mock_obs}.\footnote{Here our setup is identical except that our azimuthal ($\phi$) domain covers the whole $2\pi$ while \citetalias{XK21b} used $\phi\in[0,\pi]$ with periodic boundaries to reduce numerical cost.}
The disk is gravitationally unstable and shows prominent spirals with order-unity amplitude in the column-density profile.
However, since only the surface of the disk is visible, the observed flux is not directly proportional to the column density.
As demonstrated in Fig. \ref{fig:mock_obs}, the large-scale spirals are barely visible and we only see less coherent, small-scale perturbations at the surface of the disk, which are likely associated with shocks in the cascade of turbulent perturbations driven by the large-scale spirals.
Meanwhile, gravitationally unstable protoplanetary (Class II) disks generally suffer less from this problem, as they are cooler and have lower column density \citep[e.g.,][]{Rowther2020}.



The problem of high optical depth is less severe for wavelengths $\gtrsim$~cm, as typical protostellar disks will be optically thin at these longer wavelengths (Fig. \ref{fig:radial_profile}). However, at such long wavelengths, detection sensitivity becomes the bottleneck.
For example, for the 9-mm observations in \citetalias{Tobin2020}, only a small (and often unresolved) central portion of the disk is above the detection limit. In the future, observations with higher sensitivity and resolution at $\sim$~cm wavelength might be a direct probe for spiral structures in protostellar disks.

\begin{figure}
    \centering
    \includegraphics[scale=.66]{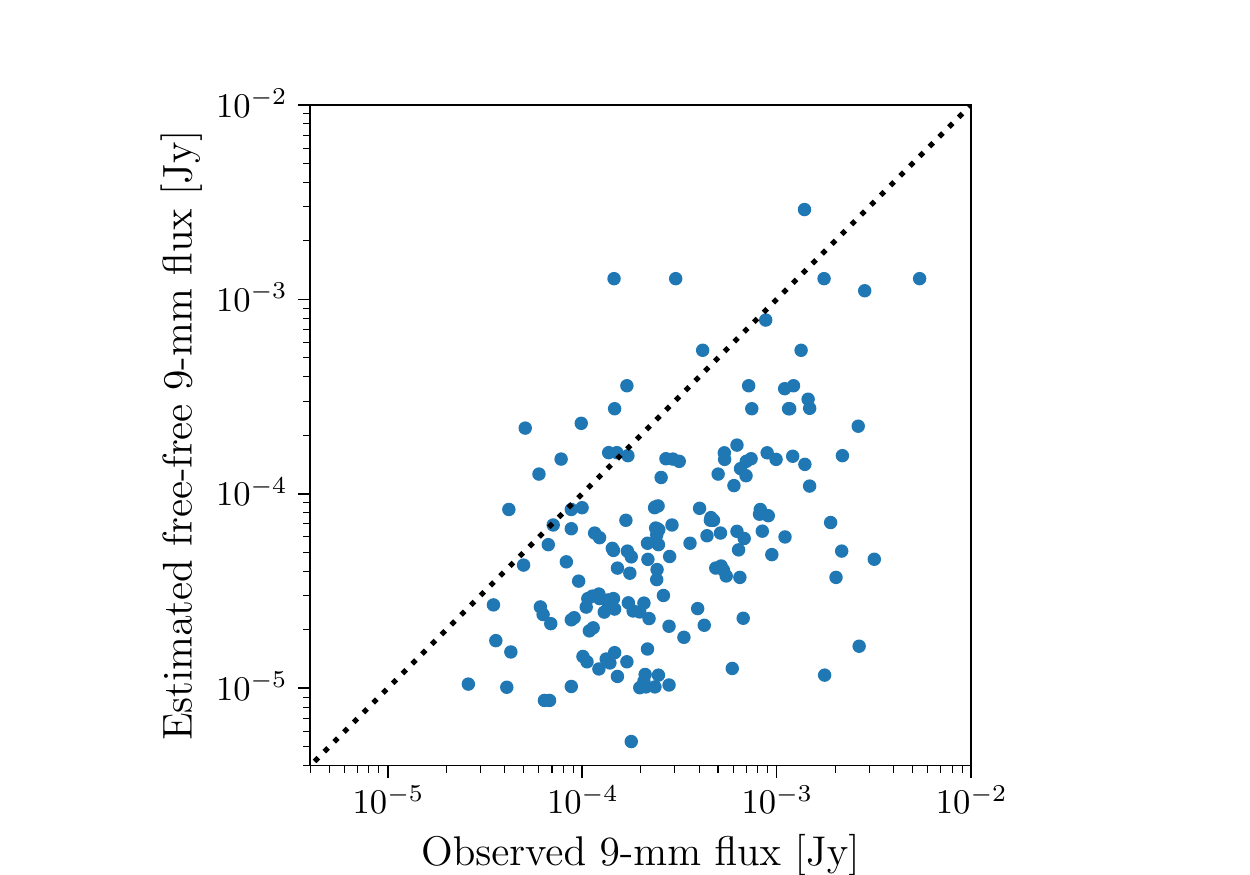}
    \includegraphics[scale=.66]{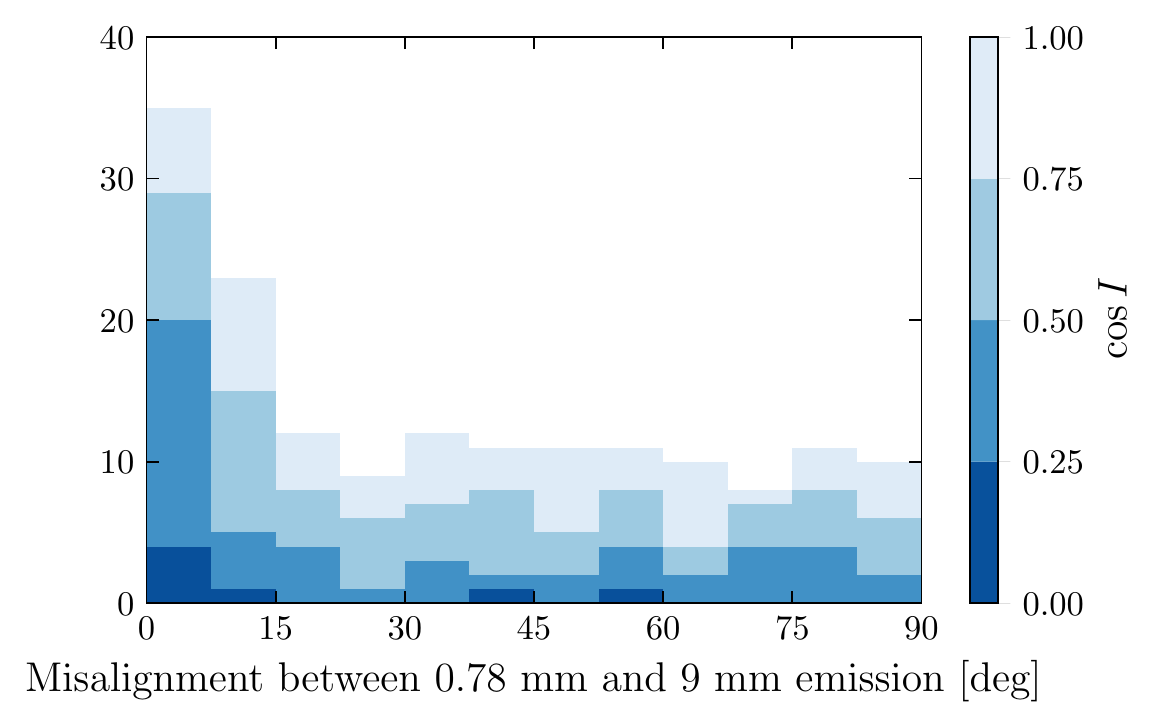}
    \caption{Top: contribution from free-free emission to 9-mm flux estimated with empirical scaling relations in \citet{Tychoniec2018}. Bottom: the distribution of misalignment between orientations of 0.87-mm and 9-mm emission from PA estimates in \citetalias{Tobin2020}, showing a peak for aligned emission and a large base of unresolved systems. Both results suggest that free-free emission in the outflow is unlikely to dominate the observed 9-mm flux.}
    \label{fig:free_free_alignment}
\end{figure}
\subsection{Is the 9-mm flux contaminated by free-free emission?}
In this paper we have assumed that flux at both observed wavelengths are dominated by dust thermal emission. While this assumption is reasonable at 0.87 mm, the 9-mm flux may be subject to contamination from free-free emission coming from ionized gas in the outflow. \citet{Tychoniec2018} studied radio emission in Perseus and concluded that ${\sim}60\%$ of the 9-mm flux there could be due to free-free emission, based on an extrapolation of (free-free dominated) fluxes at 4.1 cm and 6.4 cm. Here we evaluate whether our sample of Orion protostars also suffer from a high level of free-free contamination at 9 mm.

We begin with a rough estimate using results from \citet{Tychoniec2018}, which fitted an empirical relation between $L_{\rm bol}$ and 4.1-cm luminosity, and find that the typical spectral index of free-free emission (between 4.1~cm and 6.4~cm) is 0.3--0.4. Using this empirical relation and assuming a free-free spectral index of 0.4 between 9~mm and 4.1~cm, we can estimate the free-free luminosity at 9~mm from $L_{\rm bol}$. This estimated 9-mm free-free emission is generally lower than the observed 9-mm emission, with median $F^{\rm free-free}_{\rm 9mm}/F_{\rm 9mm} \sim 0.2$ (Fig. \ref{fig:free_free_alignment}, top panel). Therefore, the observed 9-mm flux should still be dominated by dust thermal emission.

However, this conclusion is subject to some uncertainty given the spread in the estimated 9-mm flux. It is also unclear whether the empirical scaling laws in Perseus would also be applicable to Orion. To address these caveats, we check another diagnostic which is less quantitative but more robust: the misalignment between the orientation (PA) of 0.87-mm and 9-mm emissions. When the source is resolved, the 9-mm emission should be aligned with the 0.87-mm emission if it is dominated by dust in the disk, and approximately perpendicular if it is dominated by free-free emission in the outflow. In the bottom panel of Fig. \ref{fig:free_free_alignment} we plot the distribution of the estimated misalignment; the distribution peaks at $0^\circ$ and does not show any peak around $90^\circ$, consistent with disk-dominated 9-mm emission. One caveat, however, is that most systems in our sample are unresolved or barely resolved at 9~mm; for these systems there are no accurate estimates of orientation, which results in a large ``base" of random estimated misalignment in the distribution, and we cannot directly confirm whether they have disk-dominated emission. This problem can be somewhat alleviated by looking at subsamples of low-inclination systems; for example, for systems with $\cos I\leq 0.5$ (0.25), the fraction of aligned systems ($\Delta{\rm PA}<15^\circ$) increases to 47\% (71\%).

In summary, current data prefer disk-dominated emission at 9-mm, yet future observations at longer wavelengths and/or higher resolution are needed to more reliably constrain the level of free-free contamination.

\subsection{Uncertainties in the physical picture and other scenarios of disk formation}\label{sec:other_formation_scenarios}

The physical picture of protostellar disk formation can be diverse. While in this paper we demonstrate that the observed Class 0/I disks are consistent with gravitational self-regulation at the population level, this does not rule out other scenarios of disk formation. Here we discuss the main (theoretical) uncertainties in our assumed physical picture (gravitational self-regulation), and how they result in other possible scenarios of disk formation.

One major uncertainty in our argument in Section \ref{sec:physical_assumption_Q} is whether the disk is in a quasi-steady state where accretion from the envelope onto the disk is approximately balanced by angular-momentum transport within the disk.
For example, if the initial condition of the pre-stellar core and subsequent envelope evolution are highly turbulent, the specific angular momentum of the material being accreted onto the disk might undergo large variations, and the disk evolution might be driven mainly by such variations as opposed to angular-momentum transport within the disk.
In the most extreme case, the whole disk might just be formed by a small jet of high-angular-momentum material; this is one possible explanation for the small number of systems in the VANDAM Orion survey that show prominent rings \citep{Sheehan2020}.
From a theoretical point of view, it is unclear whether the envelope should be highly turbulent and the accretion onto the disk highly variable. Some molecular-cloud-scale simulations support this possibility \citep[e.g.,][]{Kuffmeier2017,Kuznetsova2020}, with the caveat that the adoption of ideal MHD in these studies could overestimate the level of turbulence at and below core/envelope scale.

Another uncertainty in our argument is whether ambipolar diffusion inside the disk is strong enough to render magnetically driven angular-momentum transport/removal negligible. This could depend on both initial conditions and disk chemistry, which are subject to large uncertainties. Especially, the abundance of small grains plays an important role in determining the strength of non-ideal MHD effects \citep{Zhao2018}.
If the magnetic field is better coupled to the gas, disk evolution (and accretion onto the protostar) could be driven mainly by magnetic braking and wind launching and the disk could be gravitationally stable and less turbulent.
As a side note, it is also possible that this kind of magnetic regulation co-exists with gravitational self-regulation in some disks, especially during later times (e.g., Flat Spectrum) when the accretion rate drops.

\section{Summary}\label{sec:summary}
In this paper, we formulate a parametrized disk model that generate radial profiles and mock observations of embedded protostellar disks; this model can be used to infer disk properties from multi-wavelength dust-continuum observations (Sections \ref{sec:physical_assumptions}-\ref{sec:fit}).

The central assumption of our model is that the disk is gravitationally self-regulated and marginally gravitationally unstable (due to the presence of an infalling envelope). This and other physical assumptions of our model are motivated by a recent theoretical study of protostellar disk formation (\citetalias{XK21b}; Section \ref{sec:physical_assumptions}).
The adoption of these assumptions reduces the number of free parameters in our model without making arbitrary assumptions on dust temperature and disk optical depth. Especially, our model can produce reliable disk mass estimates even when the disk is optically thick.

We find that this model fits relatively well to the majority of the protostellar disks in the VANDAM Orion survey (\citetalias{Tobin2020}; Section \ref{sec:test}).
Moreover, the observations prefer our fiducial model with marginally unstable Toomre $Q$ compared to models with larger $Q$ values, suggesting that the assumption of gravitational self-regulation is likely valid at the population level.
(Note that this does not rule out other scenarios of disk formation; cf. Section \ref{sec:other_formation_scenarios})

Using our model, we produce new estimates of Orion protostellar disk properties (Section \ref{sec:results}).
Our main findings include:
\begin{itemize}
\item Disks are significantly more massive than previously expected, with typical disk-to-star mass ratio $M_{\rm d}/M_\star = \mathcal O(1)$. The high optical depth at $\lesssim$~mm wavelengths could have caused a systematic underestimation of disk mass in previous studies (e.g., \citetalias{Tobin2020}).
\item Both disk mass and disk size decrease towards later stages of protostellar evolution. These trends might be associated with the decrease in accretion rate and the pile-up of magnetic flux in the inner envelope at later times. In general, the evolution of disk properties throughout its lifetime is determined by a competition of several different mechanisms and can be non-monotonic.
\item Our model suggests that most disks in our sample are likely prone to fragmentation beyond 20-50~au, with the caveat that the applicability of our model to fragmenting disks cannot be guaranteed.
\end{itemize}

One limitation of our model is that its estimates are subject to large uncertainties, mainly due to the order-magnitude systematic uncertainties in assumed model parameters (Section \ref{sec:systematic_uncertainties}). While the qualitative trends reported in this paper, including the $\mathcal O(1)$ disk-to-star mass ratio, are robust against these uncertainties, the estimates for key disk properties are subject to uncertainties of a factor of a few. In the future, these uncertainties can be reduced if observations provide additional constraints on protostar mass, accretion rate, or grain size distribution.

Scripts for using our model to fit multi-wavelength dust-continuum images can be downloaded at \href{https://github.com/wxu26/GIdisk2obs}{https://github.com/wxu26/GIdisk2obs}. The repository also contains the estimated properties (including radial profiles) of individual systems in our sample.

\begin{acknowledgments}
We thank Matthew Kunz for insightful discussions and detailed comments on a draft version of this paper, Patrick Sheehan and Steven Stahler for discussions on comparing different methods of estimating disk properties, John Tobin for discussions on estimating the amplitude of free-free emission, and the anonymous referees for pointing out the importance of scattering opacity and improving the presentation of the paper.
The simulation presented in Fig. \ref{fig:mock_obs} was performed on computational resources managed and supported by Princeton Research Computing, a consortium of groups including the Princeton Institute for Computational Science and Engineering (PICSciE) and the Office of Information Technology's High Performance Computing Center and Visualization Laboratory at Princeton University.
\end{acknowledgments}

\vspace{5mm}
\software{Astropy \citep{astropy:2013, astropy:2018},
DSHARP-opac \citep{Birnstiel2018},
Matplotlib \citep{matplotlib}
}

\bibliography{Xu22}{}

\begin{thebibliography}{}
\expandafter\ifx\csname natexlab\endcsname\relax\def\natexlab#1{#1}\fi
\providecommand{\url}[1]{\href{#1}{#1}}
\providecommand{\dodoi}[1]{doi:~\href{http://doi.org/#1}{\nolinkurl{#1}}}
\providecommand{\doeprint}[1]{\href{http://ascl.net/#1}{\nolinkurl{http://ascl.net/#1}}}
\providecommand{\doarXiv}[1]{\href{https://arxiv.org/abs/#1}{\nolinkurl{https://arxiv.org/abs/#1}}}

\bibitem[{{Anderson} {et~al.}(2022){Anderson}, {Cleeves}, {Blake}, {Bergin},
  {Zhang}, {Carpenter}, \& {Schwarz}}]{Anderson2022}
{Anderson}, D.~E., {Cleeves}, L.~I., {Blake}, G.~A., {et~al.} 2022, arXiv
  e-prints, arXiv:2202.00709.
\newblock \doarXiv{2202.00709}

\bibitem[{{Andre} {et~al.}(2000){Andre}, {Ward-Thompson}, \&
  {Barsony}}]{Andre2000}
{Andre}, P., {Ward-Thompson}, D., \& {Barsony}, M. 2000, in Protostars and
  Planets IV, ed. V.~{Mannings}, A.~P. {Boss}, \& S.~S. {Russell}, 59.
\newblock \doarXiv{astro-ph/9903284}

\bibitem[{{Andrews} {et~al.}(2013){Andrews}, {Rosenfeld}, {Kraus}, \&
  {Wilner}}]{Andrews2013}
{Andrews}, S.~M., {Rosenfeld}, K.~A., {Kraus}, A.~L., \& {Wilner}, D.~J. 2013,
  \apj, 771, 129, \dodoi{10.1088/0004-637X/771/2/129}

\bibitem[{{Ansdell} {et~al.}(2016){Ansdell}, {Williams}, {van der Marel},
  {Carpenter}, {Guidi}, {Hogerheijde}, {Mathews}, {Manara}, {Miotello},
  {Natta}, {Oliveira}, {Tazzari}, {Testi}, {van Dishoeck}, \& {van
  Terwisga}}]{Ansdell2016}
{Ansdell}, M., {Williams}, J.~P., {van der Marel}, N., {et~al.} 2016, \apj,
  828, 46, \dodoi{10.3847/0004-637X/828/1/46}

\bibitem[{{Ansdell} {et~al.}(2018){Ansdell}, {Williams}, {Trapman}, {van
  Terwisga}, {Facchini}, {Manara}, {van der Marel}, {Miotello}, {Tazzari},
  {Hogerheijde}, {Guidi}, {Testi}, \& {van Dishoeck}}]{Ansdell2018}
{Ansdell}, M., {Williams}, J.~P., {Trapman}, L., {et~al.} 2018, \apj, 859, 21,
  \dodoi{10.3847/1538-4357/aab890}

\bibitem[{{Astropy Collaboration} {et~al.}(2013){Astropy Collaboration},
  {Robitaille}, {Tollerud}, {Greenfield}, {Droettboom}, {Bray}, {Aldcroft},
  {Davis}, {Ginsburg}, {Price-Whelan}, {Kerzendorf}, {Conley}, {Crighton},
  {Barbary}, {Muna}, {Ferguson}, {Grollier}, {Parikh}, {Nair}, {Unther},
  {Deil}, {Woillez}, {Conseil}, {Kramer}, {Turner}, {Singer}, {Fox}, {Weaver},
  {Zabalza}, {Edwards}, {Azalee Bostroem}, {Burke}, {Casey}, {Crawford},
  {Dencheva}, {Ely}, {Jenness}, {Labrie}, {Lim}, {Pierfederici}, {Pontzen},
  {Ptak}, {Refsdal}, {Servillat}, \& {Streicher}}]{astropy:2013}
{Astropy Collaboration}, {Robitaille}, T.~P., {Tollerud}, E.~J., {et~al.} 2013,
  \aap, 558, A33, \dodoi{10.1051/0004-6361/201322068}

\bibitem[{{Astropy Collaboration} {et~al.}(2018){Astropy Collaboration},
  {Price-Whelan}, {Sip{\H{o}}cz}, {G{\"u}nther}, {Lim}, {Crawford}, {Conseil},
  {Shupe}, {Craig}, {Dencheva}, {Ginsburg}, {VanderPlas}, {Bradley},
  {P{\'e}rez-Su{\'a}rez}, {de Val-Borro}, {Aldcroft}, {Cruz}, {Robitaille},
  {Tollerud}, {Ardelean}, {Babej}, {Bach}, {Bachetti}, {Bakanov}, {Bamford},
  {Barentsen}, {Barmby}, {Baumbach}, {Berry}, {Biscani}, {Boquien}, {Bostroem},
  {Bouma}, {Brammer}, {Bray}, {Breytenbach}, {Buddelmeijer}, {Burke},
  {Calderone}, {Cano Rodr{\'\i}guez}, {Cara}, {Cardoso}, {Cheedella}, {Copin},
  {Corrales}, {Crichton}, {D'Avella}, {Deil}, {Depagne}, {Dietrich}, {Donath},
  {Droettboom}, {Earl}, {Erben}, {Fabbro}, {Ferreira}, {Finethy}, {Fox},
  {Garrison}, {Gibbons}, {Goldstein}, {Gommers}, {Greco}, {Greenfield},
  {Groener}, {Grollier}, {Hagen}, {Hirst}, {Homeier}, {Horton}, {Hosseinzadeh},
  {Hu}, {Hunkeler}, {Ivezi{\'c}}, {Jain}, {Jenness}, {Kanarek}, {Kendrew},
  {Kern}, {Kerzendorf}, {Khvalko}, {King}, {Kirkby}, {Kulkarni}, {Kumar},
  {Lee}, {Lenz}, {Littlefair}, {Ma}, {Macleod}, {Mastropietro}, {McCully},
  {Montagnac}, {Morris}, {Mueller}, {Mumford}, {Muna}, {Murphy}, {Nelson},
  {Nguyen}, {Ninan}, {N{\"o}the}, {Ogaz}, {Oh}, {Parejko}, {Parley}, {Pascual},
  {Patil}, {Patil}, {Plunkett}, {Prochaska}, {Rastogi}, {Reddy Janga},
  {Sabater}, {Sakurikar}, {Seifert}, {Sherbert}, {Sherwood-Taylor}, {Shih},
  {Sick}, {Silbiger}, {Singanamalla}, {Singer}, {Sladen}, {Sooley},
  {Sornarajah}, {Streicher}, {Teuben}, {Thomas}, {Tremblay}, {Turner},
  {Terr{\'o}n}, {van Kerkwijk}, {de la Vega}, {Watkins}, {Weaver}, {Whitmore},
  {Woillez}, {Zabalza}, \& {Astropy Contributors}}]{astropy:2018}
{Astropy Collaboration}, {Price-Whelan}, A.~M., {Sip{\H{o}}cz}, B.~M., {et~al.}
  2018, \aj, 156, 123, \dodoi{10.3847/1538-3881/aabc4f}

\bibitem[{{Baehr} {et~al.}(2022){Baehr}, {Zhu}, \& {Yang}}]{Baehr2022}
{Baehr}, H., {Zhu}, Z., \& {Yang}, C.-C. 2022, arXiv e-prints,
  arXiv:2204.13310.
\newblock \doarXiv{2204.13310}

\bibitem[{{Balbus} {et~al.}(1994){Balbus}, {Gammie}, \& {Hawley}}]{bgh94}
{Balbus}, S.~A., {Gammie}, C.~F., \& {Hawley}, J.~F. 1994, \mnras, 271, 197,
  \dodoi{10.1093/mnras/271.1.197}

\bibitem[{{Bate}(2018)}]{Bate2018}
{Bate}, M.~R. 2018, \mnras, 475, 5618, \dodoi{10.1093/mnras/sty169}

\bibitem[{{Birnstiel} {et~al.}(2010){Birnstiel}, {Dullemond}, \&
  {Brauer}}]{Birnstiel2010}
{Birnstiel}, T., {Dullemond}, C.~P., \& {Brauer}, F. 2010, \aap, 513, A79,
  \dodoi{10.1051/0004-6361/200913731}

\bibitem[{{Birnstiel} {et~al.}(2011){Birnstiel}, {Ormel}, \&
  {Dullemond}}]{Birnstiel2011}
{Birnstiel}, T., {Ormel}, C.~W., \& {Dullemond}, C.~P. 2011, \aap, 525, A11,
  \dodoi{10.1051/0004-6361/201015228}

\bibitem[{{Birnstiel} {et~al.}(2018){Birnstiel}, {Dullemond}, {Zhu}, {Andrews},
  {Bai}, {Wilner}, {Carpenter}, {Huang}, {Isella}, {Benisty}, {P{\'e}rez}, \&
  {Zhang}}]{Birnstiel2018}
{Birnstiel}, T., {Dullemond}, C.~P., {Zhu}, Z., {et~al.} 2018, \apjl, 869, L45,
  \dodoi{10.3847/2041-8213/aaf743}

\bibitem[{{Booth} {et~al.}(2019){Booth}, {Walsh}, {Ilee}, {Notsu}, {Qi},
  {Nomura}, \& {Akiyama}}]{Booth2019}
{Booth}, A.~S., {Walsh}, C., {Ilee}, J.~D., {et~al.} 2019, \apjl, 882, L31,
  \dodoi{10.3847/2041-8213/ab3645}

\bibitem[{{Clarke}(2009)}]{Clarke2009}
{Clarke}, C.~J. 2009, \mnras, 396, 1066,
  \dodoi{10.1111/j.1365-2966.2009.14774.x}

\bibitem[{{D'Alessio} {et~al.}(1997){D'Alessio}, {Calvet}, \&
  {Hartmann}}]{DAlessio1997}
{D'Alessio}, P., {Calvet}, N., \& {Hartmann}, L. 1997, \apj, 474, 397,
  \dodoi{10.1086/303433}

\bibitem[{{Draine}(2006)}]{Draine2006}
{Draine}, B.~T. 2006, \apj, 636, 1114, \dodoi{10.1086/498130}

\bibitem[{{Eisner} {et~al.}(2018){Eisner}, {Arce}, {Ballering}, {Bally},
  {Andrews}, {Boyden}, {Di Francesco}, {Fang}, {Johnstone}, {Kim}, {Mann},
  {Matthews}, {Pascucci}, {Ricci}, {Sheehan}, \& {Williams}}]{Eisner2018}
{Eisner}, J.~A., {Arce}, H.~G., {Ballering}, N.~P., {et~al.} 2018, \apj, 860,
  77, \dodoi{10.3847/1538-4357/aac3e2}

\bibitem[{{Fiedler} \& {Mouschovias}(1993)}]{FiedlerMouschovias1993}
{Fiedler}, R.~A., \& {Mouschovias}, T.~C. 1993, \apj, 415, 680,
  \dodoi{10.1086/173193}

\bibitem[{{Fischer} {et~al.}(2017){Fischer}, {Megeath}, {Furlan}, {Ali},
  {Stutz}, {Tobin}, {Osorio}, {Stanke}, {Manoj}, {Poteet}, {Booker},
  {Hartmann}, {Wilson}, {Myers}, \& {Watson}}]{Fischer2017}
{Fischer}, W.~J., {Megeath}, S.~T., {Furlan}, E., {et~al.} 2017, \apj, 840, 69,
  \dodoi{10.3847/1538-4357/aa6d69}

\bibitem[{{Galli} \& {Shu}(1993)}]{GalliShu1993}
{Galli}, D., \& {Shu}, F.~H. 1993, \apj, 417, 220, \dodoi{10.1086/173305}

\bibitem[{{Galv{\'a}n-Madrid} {et~al.}(2018){Galv{\'a}n-Madrid}, {Liu},
  {Izquierdo}, {Miotello}, {Zhao}, {Carrasco-Gonz{\'a}lez}, {Lizano}, \&
  {Rodr{\'\i}guez}}]{Galvan-Madrid2018}
{Galv{\'a}n-Madrid}, R., {Liu}, H.~B., {Izquierdo}, A.~F., {et~al.} 2018, \apj,
  868, 39, \dodoi{10.3847/1538-4357/aae779}

\bibitem[{{Gammie}(1996)}]{Gammie1996}
{Gammie}, C.~F. 1996, \apj, 457, 355, \dodoi{10.1086/176735}

\bibitem[{{Gammie}(2001)}]{Gammie2001}
---. 2001, \apj, 553, 174, \dodoi{10.1086/320631}

\bibitem[{{Hubeny}(1990)}]{Hubeny1990}
{Hubeny}, I. 1990, \apj, 351, 632, \dodoi{10.1086/168501}

\bibitem[{Hunter(2007)}]{matplotlib}
Hunter, J.~D. 2007, Computing in Science Engineering, 9, 90,
  \dodoi{10.1109/MCSE.2007.55}

\bibitem[{{Ishimaru}(1978)}]{Ishimaru1978}
{Ishimaru}, A. 1978, {Wave propagation and scattering in random media. Volume 1
  - Single scattering and transport theory}, Vol.~1,
  \dodoi{10.1016/B978-0-12-374701-3.X5001-7}

\bibitem[{{Kratter} \& {Lodato}(2016)}]{KratterLodato2016}
{Kratter}, K., \& {Lodato}, G. 2016, \araa, 54, 271,
  \dodoi{10.1146/annurev-astro-081915-023307}

\bibitem[{{Kristensen} {et~al.}(2012){Kristensen}, {van Dishoeck}, {Bergin},
  {Visser}, {Y{\i}ld{\i}z}, {San Jose-Garcia}, {J{\o}rgensen}, {Herczeg},
  {Johnstone}, {Wampfler}, {Benz}, {Bruderer}, {Cabrit}, {Caselli}, {Doty},
  {Harsono}, {Herpin}, {Hogerheijde}, {Karska}, {van Kempen}, {Liseau},
  {Nisini}, {Tafalla}, {van der Tak}, \& {Wyrowski}}]{Kristensen2012}
{Kristensen}, L.~E., {van Dishoeck}, E.~F., {Bergin}, E.~A., {et~al.} 2012,
  \aap, 542, A8, \dodoi{10.1051/0004-6361/201118146}

\bibitem[{{Kuffmeier} {et~al.}(2017){Kuffmeier}, {Haugb{\o}lle}, \&
  {Nordlund}}]{Kuffmeier2017}
{Kuffmeier}, M., {Haugb{\o}lle}, T., \& {Nordlund}, {\r{A}}. 2017, \apj, 846,
  7, \dodoi{10.3847/1538-4357/aa7c64}

\bibitem[{{Kuznetsova} {et~al.}(2020){Kuznetsova}, {Hartmann}, \&
  {Heitsch}}]{Kuznetsova2020}
{Kuznetsova}, A., {Hartmann}, L., \& {Heitsch}, F. 2020, \apj, 893, 73,
  \dodoi{10.3847/1538-4357/ab7eac}

\bibitem[{{Lebreuilly} {et~al.}(2020){Lebreuilly}, {Commer{\c{c}}on}, \&
  {Laibe}}]{Lebreuilly2020}
{Lebreuilly}, U., {Commer{\c{c}}on}, B., \& {Laibe}, G. 2020, \aap, 641, A112,
  \dodoi{10.1051/0004-6361/202038174}

\bibitem[{{Li} {et~al.}(2017){Li}, {Liu}, {Hasegawa}, \& {Hirano}}]{Li2017}
{Li}, J. I.-H., {Liu}, H.~B., {Hasegawa}, Y., \& {Hirano}, N. 2017, \apj, 840,
  72, \dodoi{10.3847/1538-4357/aa6f04}

\bibitem[{{Li} \& {McKee}(1996)}]{LiMcKee1996}
{Li}, Z.-Y., \& {McKee}, C.~F. 1996, \apj, 464, 373, \dodoi{10.1086/177329}

\bibitem[{{Lin} \& {Pringle}(1990)}]{LinPringle1990}
{Lin}, D. N.~C., \& {Pringle}, J.~E. 1990, \apj, 358, 515,
  \dodoi{10.1086/169004}

\bibitem[{{Masson} {et~al.}(2016){Masson}, {Chabrier}, {Hennebelle}, {Vaytet},
  \& {Commer{\c{c}}on}}]{Masson2016}
{Masson}, J., {Chabrier}, G., {Hennebelle}, P., {Vaytet}, N., \&
  {Commer{\c{c}}on}, B. 2016, \aap, 587, A32,
  \dodoi{10.1051/0004-6361/201526371}

\bibitem[{{Mathis} {et~al.}(1977){Mathis}, {Rumpl}, \&
  {Nordsieck}}]{Mathis1977}
{Mathis}, J.~S., {Rumpl}, W., \& {Nordsieck}, K.~H. 1977, \apj, 217, 425,
  \dodoi{10.1086/155591}

\bibitem[{{Miyake} \& {Nakagawa}(1993)}]{MiyakeNakagawa1993}
{Miyake}, K., \& {Nakagawa}, Y. 1993, \icarus, 106, 20,
  \dodoi{10.1006/icar.1993.1156}

\bibitem[{{Natta}(1993)}]{Natta1993}
{Natta}, A. 1993, \apj, 412, 761, \dodoi{10.1086/172959}

\bibitem[{{Offner} \& {McKee}(2011)}]{OffnerMcKee2011}
{Offner}, S. S.~R., \& {McKee}, C.~F. 2011, \apj, 736, 53,
  \dodoi{10.1088/0004-637X/736/1/53}

\bibitem[{{Paczynski}(1978)}]{Paczynski1978}
{Paczynski}, B. 1978, \actaa, 28, 91

\bibitem[{{Pineda} {et~al.}(2012){Pineda}, {Maury}, {Fuller}, {Testi},
  {Garc{\'\i}a-Appadoo}, {Peck}, {Villard}, {Corder}, {van Kempen}, {Turner},
  {Tachihara}, \& {Dent}}]{Pineda2012}
{Pineda}, J.~E., {Maury}, A.~J., {Fuller}, G.~A., {et~al.} 2012, \aap, 544, L7,
  \dodoi{10.1051/0004-6361/201219589}

\bibitem[{{Pollack} {et~al.}(1994){Pollack}, {Hollenbach}, {Beckwith},
  {Simonelli}, {Roush}, \& {Fong}}]{Pollack1994}
{Pollack}, J.~B., {Hollenbach}, D., {Beckwith}, S., {et~al.} 1994, \apj, 421,
  615, \dodoi{10.1086/173677}

\bibitem[{{Powell} {et~al.}(2019){Powell}, {Murray-Clay}, {P{\'e}rez},
  {Schlichting}, \& {Rosenthal}}]{Powell2019}
{Powell}, D., {Murray-Clay}, R., {P{\'e}rez}, L.~M., {Schlichting}, H.~E., \&
  {Rosenthal}, M. 2019, \apj, 878, 116, \dodoi{10.3847/1538-4357/ab20ce}

\bibitem[{{Rafikov}(2007)}]{Rafikov2007}
{Rafikov}, R.~R. 2007, \apj, 662, 642, \dodoi{10.1086/517599}

\bibitem[{{Rafikov}(2009)}]{Rafikov2009}
---. 2009, \apj, 704, 281, \dodoi{10.1088/0004-637X/704/1/281}

\bibitem[{{Rowther} {et~al.}(2020){Rowther}, {Meru}, {Kennedy}, {Nealon}, \&
  {Pinte}}]{Rowther2020}
{Rowther}, S., {Meru}, F., {Kennedy}, G.~M., {Nealon}, R., \& {Pinte}, C. 2020,
  \apjl, 904, L18, \dodoi{10.3847/2041-8213/abc704}

\bibitem[{{Segura-Cox} {et~al.}(2018){Segura-Cox}, {Looney}, {Tobin}, {Li},
  {Harris}, {Sadavoy}, {Dunham}, {Chandler}, {Kratter}, {P{\'e}rez}, \&
  {Melis}}]{Segura-Cox2018}
{Segura-Cox}, D.~M., {Looney}, L.~W., {Tobin}, J.~J., {et~al.} 2018, \apj, 866,
  161, \dodoi{10.3847/1538-4357/aaddf3}

\bibitem[{{Sheehan} \& {Eisner}(2017)}]{SheehanEisner2017}
{Sheehan}, P.~D., \& {Eisner}, J.~A. 2017, \apj, 851, 45,
  \dodoi{10.3847/1538-4357/aa9990}

\bibitem[{{Sheehan} {et~al.}(2020){Sheehan}, {Tobin}, {Federman}, {Megeath}, \&
  {Looney}}]{Sheehan2020}
{Sheehan}, P.~D., {Tobin}, J.~J., {Federman}, S., {Megeath}, S.~T., \&
  {Looney}, L.~W. 2020, \apj, 902, 141, \dodoi{10.3847/1538-4357/abbad5}

\bibitem[{{Sheehan} {et~al.}(2022){Sheehan}, {Tobin}, {Looney}, \&
  {Megeath}}]{Sheehan2022}
{Sheehan}, P.~D., {Tobin}, J.~J., {Looney}, L.~L., \& {Megeath}, S.~T. 2022,
  arXiv e-prints, arXiv:2203.00029.
\newblock \doarXiv{2203.00029}

\bibitem[{{Tassis} \& {Mouschovias}(2005)}]{TassisMouschovias2005}
{Tassis}, K., \& {Mouschovias}, T.~C. 2005, \apj, 618, 783,
  \dodoi{10.1086/424480}

\bibitem[{Tobin(2019{\natexlab{a}})}]{VANDAMOrionALMA}
Tobin, J. 2019{\natexlab{a}}, {ALMA 870 micron Continuum Measurement Sets}, V1,
   Harvard Dataverse, \dodoi{10.7910/DVN/5NXMLI}

\bibitem[{Tobin(2019{\natexlab{b}})}]{VANDAMOrionVLA}
---. 2019{\natexlab{b}}, {VLA Ka-band (9 mm) Continuum - A-configuration (0.08
  arcsec)}, V1,  Harvard Dataverse, \dodoi{10.7910/DVN/ICKYX0}

\bibitem[{{Tobin} {et~al.}(2016){Tobin}, {Looney}, {Li}, {Chandler}, {Dunham},
  {Segura-Cox}, {Sadavoy}, {Melis}, {Harris}, {Kratter}, \&
  {Perez}}]{Tobin2016}
{Tobin}, J.~J., {Looney}, L.~W., {Li}, Z.-Y., {et~al.} 2016, \apj, 818, 73,
  \dodoi{10.3847/0004-637X/818/1/73}

\bibitem[{{Tobin} {et~al.}(2020){Tobin}, {Sheehan}, {Megeath},
  {D{\'\i}az-Rodr{\'\i}guez}, {Offner}, {Murillo}, {van 't Hoff}, {van
  Dishoeck}, {Osorio}, {Anglada}, {Furlan}, {Stutz}, {Reynolds}, {Karnath},
  {Fischer}, {Persson}, {Looney}, {Li}, {Stephens}, {Chand ler}, {Cox},
  {Dunham}, {Tychoniec}, {Kama}, {Kratter}, {Kounkel}, {Mazur}, {Maud},
  {Patel}, {Perez}, {Sadavoy}, {Segura-Cox}, {Sharma}, {Stephenson}, {Watson},
  \& {Wyrowski}}]{Tobin2020}
{Tobin}, J.~J., {Sheehan}, P.~D., {Megeath}, S.~T., {et~al.} 2020, \apj, 890,
  130, \dodoi{10.3847/1538-4357/ab6f64}

\bibitem[{{Tobin} {et~al.}(2022){Tobin}, {Offner}, {Kratter}, {Megeath},
  {Sheehan}, {Looney}, {Diaz-Rodriguez}, {Osorio}, {Anglada}, {Sadavoy},
  {Furlan}, {Segura-Cox}, {Karnath}, {van't Hoff}, {van Dishoeck}, {Li},
  {Sharma}, {Stutz}, \& {Tychoniec}}]{Tobin2022}
{Tobin}, J.~J., {Offner}, S. S.~R., {Kratter}, K.~M., {et~al.} 2022, \apj, 925,
  39, \dodoi{10.3847/1538-4357/ac36d2}

\bibitem[{{Tripathi} {et~al.}(2017){Tripathi}, {Andrews}, {Birnstiel}, \&
  {Wilner}}]{Tripathi2017}
{Tripathi}, A., {Andrews}, S.~M., {Birnstiel}, T., \& {Wilner}, D.~J. 2017,
  \apj, 845, 44, \dodoi{10.3847/1538-4357/aa7c62}

\bibitem[{{Tychoniec} {et~al.}(2018){Tychoniec}, {Tobin}, {Karska}, {Chandler},
  {Dunham}, {Harris}, {Kratter}, {Li}, {Looney}, {Melis}, {P{\'e}rez},
  {Sadavoy}, {Segura-Cox}, \& {van Dishoeck}}]{Tychoniec2018}
{Tychoniec}, {\L}., {Tobin}, J.~J., {Karska}, A., {et~al.} 2018, \apjs, 238,
  19, \dodoi{10.3847/1538-4365/aaceae}

\bibitem[{{Umebayashi}(1983)}]{Umebayashi1983}
{Umebayashi}, T. 1983, Progress of Theoretical Physics, 69, 480,
  \dodoi{10.1143/PTP.69.480}

\bibitem[{{Vorobyov} \& {Basu}(2006)}]{VB06}
{Vorobyov}, E.~I., \& {Basu}, S. 2006, \apj, 650, 956, \dodoi{10.1086/507320}

\bibitem[{{Vorobyov} \& {Basu}(2007)}]{VB07}
---. 2007, \mnras, 381, 1009, \dodoi{10.1111/j.1365-2966.2007.12321.x}

\bibitem[{{Vorobyov} {et~al.}(2013){Vorobyov}, {DeSouza}, \& {Basu}}]{VB13}
{Vorobyov}, E.~I., {DeSouza}, A.~L., \& {Basu}, S. 2013, \apj, 768, 131,
  \dodoi{10.1088/0004-637X/768/2/131}

\bibitem[{{Vorobyov} \& {Elbakyan}(2019)}]{Vorobyov2019}
{Vorobyov}, E.~I., \& {Elbakyan}, V.~G. 2019, \aap, 631, A1,
  \dodoi{10.1051/0004-6361/201936132}

\bibitem[{{Williams} {et~al.}(2019){Williams}, {Cieza}, {Hales}, {Ansdell},
  {Ruiz-Rodriguez}, {Casassus}, {Perez}, \& {Zurlo}}]{Williams2019}
{Williams}, J.~P., {Cieza}, L., {Hales}, A., {et~al.} 2019, \apjl, 875, L9,
  \dodoi{10.3847/2041-8213/ab1338}

\bibitem[{{Woitke} {et~al.}(2016){Woitke}, {Min}, {Pinte}, {Thi}, {Kamp},
  {Rab}, {Anthonioz}, {Antonellini}, {Baldovin-Saavedra}, {Carmona}, {Dominik},
  {Dionatos}, {Greaves}, {G{\"u}del}, {Ilee}, {Liebhart}, {M{\'e}nard},
  {Rigon}, {Waters}, {Aresu}, {Meijerink}, \& {Spaans}}]{Woitke2016}
{Woitke}, P., {Min}, M., {Pinte}, C., {et~al.} 2016, \aap, 586, A103,
  \dodoi{10.1051/0004-6361/201526538}

\bibitem[{{Xu} \& {Kunz}(2021{\natexlab{a}})}]{XK21a}
{Xu}, W., \& {Kunz}, M.~W. 2021{\natexlab{a}}, \mnras, 502, 4911,
  \dodoi{10.1093/mnras/stab314}

\bibitem[{{Xu} \& {Kunz}(2021{\natexlab{b}})}]{XK21b}
---. 2021{\natexlab{b}}, \mnras, 508, 2142, \dodoi{10.1093/mnras/stab2715}

\bibitem[{{Zakri} {et~al.}(2022){Zakri}, {Megeath}, {Fischer}, {Gutermuth},
  {Furlan}, {Hartmann}, {Karnath}, {Osorio}, {Safron}, {Stanke}, {Stutz},
  {Tobin}, {Allen}, {Federman}, {Habel}, {Manoj}, {Narang}, {Pokhrel},
  {Rebull}, {Sheehan}, \& {Watson}}]{Zakri2022}
{Zakri}, W., {Megeath}, S.~T., {Fischer}, W.~J., {et~al.} 2022, \apjl, 924,
  L23, \dodoi{10.3847/2041-8213/ac46ae}

\bibitem[{{Zhao} {et~al.}(2018){Zhao}, {Caselli}, {Li}, \&
  {Krasnopolsky}}]{Zhao2018}
{Zhao}, B., {Caselli}, P., {Li}, Z.-Y., \& {Krasnopolsky}, R. 2018, \mnras,
  473, 4868, \dodoi{10.1093/mnras/stx2617}

\bibitem[{{Zhao} {et~al.}(2020){Zhao}, {Tomida}, {Hennebelle}, {Tobin},
  {Maury}, {Hirota}, {S{\'a}nchez-Monge}, {Kuiper}, {Rosen}, {Bhandare},
  {Padovani}, \& {Lee}}]{Zhao2020}
{Zhao}, B., {Tomida}, K., {Hennebelle}, P., {et~al.} 2020, \ssr, 216, 43,
  \dodoi{10.1007/s11214-020-00664-z}

\bibitem[{Zhu {et~al.}(2019)Zhu, Zhang, Jiang, Kataoka, Birnstiel, Dullemond,
  Andrews, Huang, P{\'e}rez, Carpenter, {et~al.}}]{Zhu2019}
Zhu, Z., Zhang, S., Jiang, Y.-F., {et~al.} 2019, The Astrophysical Journal
  Letters, 877, L18

\end{thebibliography}
\bibliographystyle{aasjournal}
\appendix

\section{Are dust grains well-coupled to gas?}\label{a:grain_coupling}

\begin{figure*}
    \centering
    \includegraphics[scale=0.66]{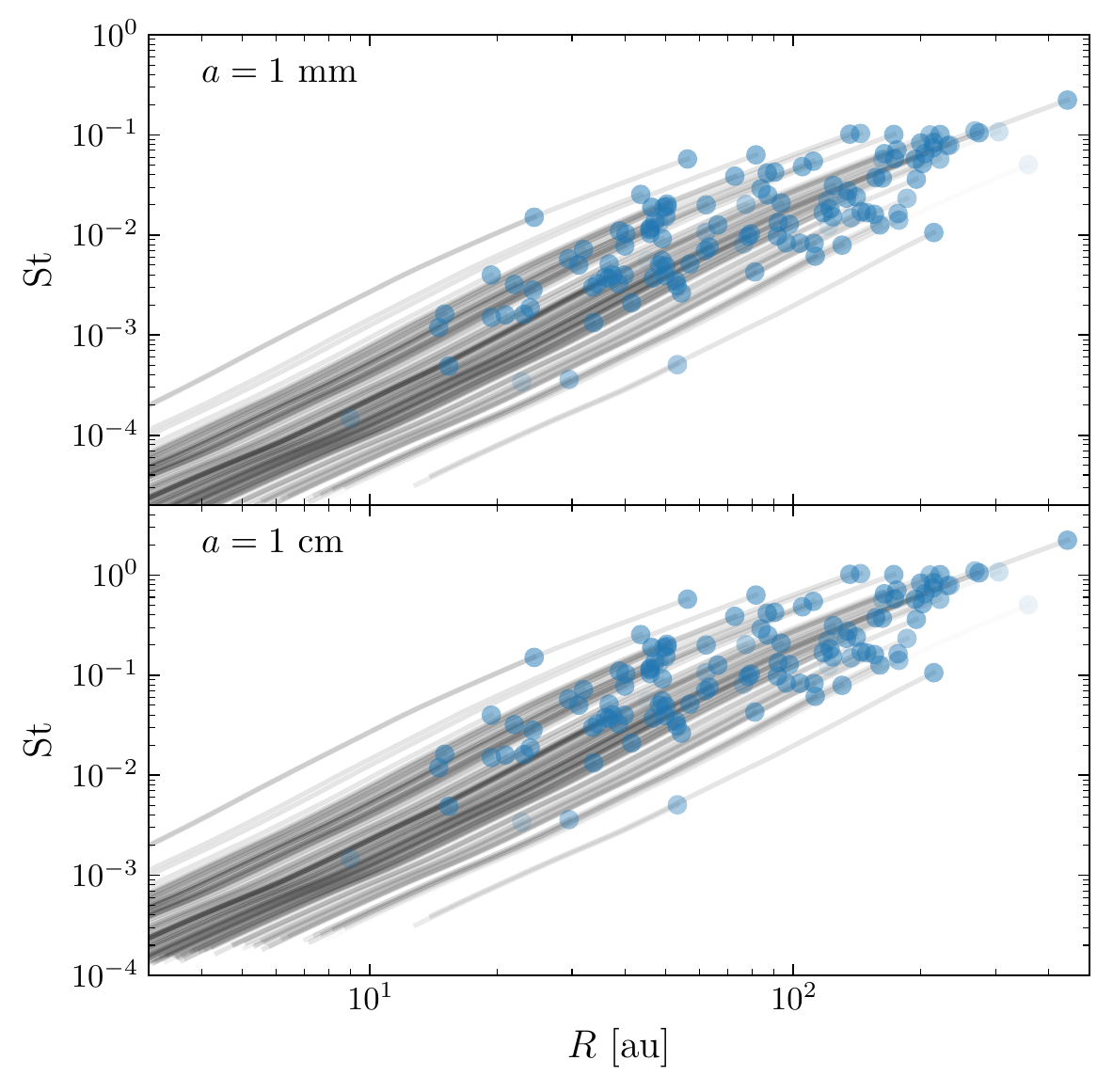} \hspace{3em} \includegraphics[scale=0.66]{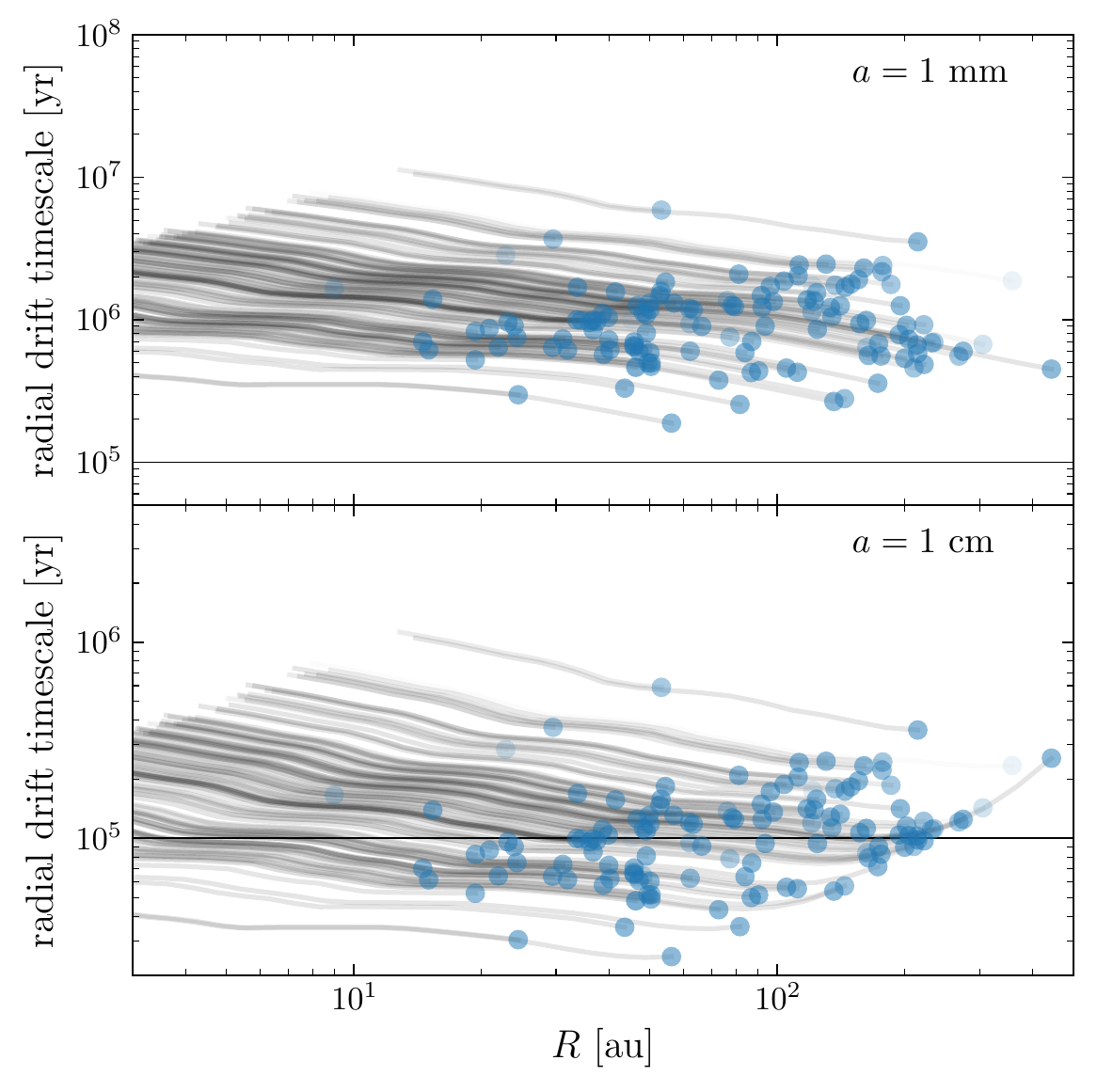}
    \caption{Left panel: Stokes number profiles for 1 mm and 1 cm grains. Right panel: Timescales of radial drift by aerodynamic drag for 1 mm and 1 cm grains. For 1 mm grains in all disks and 1 cm grains in more than half of the disks, the grains are well-coupled and radial drift by aerodynamic drag should be negligible over the lifetime of Class 0/I.}
    \label{fig:dust_drift_timescale}
\end{figure*}

In our model we have assumed a constant dust-to-gas ratio throughout the disk; here we check whether this assumption is reasonable. We start by estimating the timescale of radial drift by aerodynamic drag, which is known to cause significant dust decoupling in Class II \citep[e.g.,][]{Ansdell2018}.
Following \citet{Birnstiel2010}, we estimate the radial profile of midplane Stokes number and radial drift timescale (Fig. \ref{fig:dust_drift_timescale}) for our estimated disk profiles at two different grain sizes, 1~mm and 1~cm. We find that grains as large as ${\sim}1$~mm are still well-coupled to dust (${\rm St}\ll 1$) with a radial drift timescale longer than the typical lifetime of Class 0/I ($10^5$ yr) for all of our disks. Even for cm-size grains, less than half of our disks have sufficiently short radial timescale to allow substantial drift during Class 0/I.

There are, however, other mechanisms of decoupling that might be more important in Class 0/I. For example, the dynamical timescale in a gravitationally unstable disk could be lower than $\Omega^{-1}$ (e.g., due to shocks) and that causes the effective Stocks number to increase and makes grains less well-coupled to the gas. Recent 3D simulations by \citet{Baehr2022} shows a significant degree of clumping at St$=1$ (and much weaker clumping at St$=0.1$). More significant decoupling could occur if the disk is fragmenting (which produces shorter dynamical timescale) and dust growth in the dense fragment is modeled (which further increases St), as demonstrated in \citet{Vorobyov2019}. Magnetic field could also decouple ionized grains from the neutral gas. While this effect could be very significant in the envelope and the outflow, the resulting dust-to-gas variation in the disk, and the systematic difference between disk and pre-stellar dust-to-gas ratio, are generally a factor of a few except for very large (few cm) grains \citep{Lebreuilly2020}. In summary, while we generally should no expect a systematic difference in dust and gas disk size as we do in Class II, several mechanisms could produce enhanced and/or non-uniform dust-to-gas ratio in the disk.

\section{Systematic uncertainties and sensitivity to assumed model parameters}\label{a:uncertainty}

\begin{deluxetable*}{lccccc}
\tablecaption{Sensitivity of disk property estimates to assumed model parameters\label{tab:sensitivity}}
\startdata\\
~ & $R_{\rm d}$ & $M_{\rm d}$ & $M_\star$ & $M_{\rm d}/M_\star$ & $T_{\rm mean}$\\
\tableline
$a_{\rm max}$& $+0.03\pm0.10$ & $+0.01\pm0.05$ & $+0.04\pm0.13$ & $-0.03\pm0.11$ & $-0.04\pm0.05$ \\
$\dot M/M_\star$& $-0.03\pm0.13$ & $-0.16\pm0.06$ & $-0.45\pm0.20$ & $+0.30\pm0.22$ & $-0.01\pm0.08$ \\
$Q$& $-0.24\pm0.39$ & $-0.96\pm0.18$ & $+1.04\pm0.52$ & $-2.01\pm0.51$ & $+0.18\pm0.20$ \\
\enddata
\tablecomments{The numbers before and after $\pm$ are mean and standard deviation of $\partial \log$(disk~property)$/\partial \log$(assumed~model~parameter), respectively. These statistics are evaluated on the sub-sample with $\chi_{\rm mean}^2\leq 2$ for our fiducial model.}
\end{deluxetable*}

\begin{deluxetable}{l@{\hspace{2em}}c@{\hspace{2em}}c@{\hspace{2em}}c@{\hspace{1.5em}}c@{\hspace{1.5em}}c}
\tablecaption{Estimated $1\sigma$ uncertainties of $\log$ disk property due to each parameter\label{tab:uncertainty}}
\startdata\\
~ & $R_{\rm d}$ & $M_{\rm d}$ & $M_\star$ & $M_{\rm d}/M_\star$ & $T_{\rm mean}$\\
\tableline
$a_{\rm max}$& 0.25 & 0.11 & 0.31 & 0.27 & 0.15 \\
$\dot M/M_\star$& 0.31 & 0.39 & 1.15 & 0.85 & 0.19 \\
$Q$& 0.16 & 0.34 & 0.40 & 0.72 & 0.09 \\
\textbf{Total}& \textbf{0.43} & \textbf{0.53} & \textbf{1.25} & \textbf{1.15} & \textbf{0.26} \\
\enddata
\tablecomments{Here we assume $\sigma (\log a_{\rm max}) = \log(10)$,  $\sigma (\log \dot M/M_\star) = \log(10)$, and $\sigma (\log Q) = \log(2)/2$. We also assume that the errors in assumed model parameters are uncorrelated. These statistics are evaluated on the sub-sample with $\chi_{\rm mean}^2\leq 2$ for our fiducial model. The uncertainties in disk property estimates are between a factor of $\exp(0.26)\approx 1.3$ and $\exp(1.99)\approx 3$.}
\end{deluxetable}

In this appendix we evaluate the sensitivity of estimated disk properties on assumed parameters of our model, and estimate the resulting systematic uncertainties.

For a disk property $y$ and a set of assumed model parameter $p_i$ whose log have systematic uncertainties $\sigma(\log p_i)$, the uncertainty in $\log y$ is simply
\begin{equation}
\sigma^2(\log y) = \left\langle\sum_{i} \left(\frac{\partial \log y_j}{\partial\log p_i}\right)^2\sigma^2(\log p_i)\right\rangle_j.
\end{equation}
Here $y_j$ is the estimated value of $y$ for the $j$th system and $\langle\cdot\rangle_j$ denotes an average in $j$.
In order to estimate $\sigma(\log y)$ we first directly evaluate $\partial \log y_j/\partial\log p_i$ for each system by comparing $y_j$ fitted from different choices of $p_i$, and then assign physically reasonable $\sigma(\log p_i)$ to obtain $\sigma(\log y)$. For our model, the assumed parameters are the max dust grain size $a_{\rm max}$, the ratio between accretion rate and protostar mass $\dot M/M_\star$, and the Toomre $Q$ parameter. The first two are highly uncertain, so we assume that their $1\sigma$ uncertainties are both a factor of 10, i.e. $\sigma(\log a_{\rm max})=\sigma(\log \dot M/M_\star)=\log(10)$. For $Q$, since a gravitationally self-regulated disk generally have $Q=1-2$, we let this range correspond to $\pm 1\sigma$ and choose $\sigma(\log Q) = \log(2)/2$.

In Table \ref{tab:sensitivity} and \ref{tab:uncertainty} we summarize the evaluated $\partial \log y/\partial\log p_i$ and the estimated uncertainties of log disk properties $\sigma (\log y)$. The resulting $1\sigma$ uncertainties in disk properties range from a factor of 2-7. Therefore, disk properties estimated with our current model should generally be considered as a rough estimate. In the future, these uncertainties can be significantly reduced if data (or theory) can constrain the grain size and the accretion rate (or protostellar mass) better.

Since different disk properties are affected by the same set of parameters, their systematic errors are generally correlated. To capture this correlation, we evaluate the covariance between the systematic error of two different disk properties $(y,y')$ using
\begin{equation}
{\rm cov}(\log y, \log y') = \left\langle\sum_{i} \frac{\partial \log y_j}{\partial\log p_i}\frac{\partial \log y'_j}{\partial\log p_i}\sigma^2(\log p_i)\right\rangle_j.
\end{equation}
This covariance is used for plotting the 2D uncertainties in Figs. \ref{fig:scatter_Rd_Md} and \ref{fig:scatter_Ms_Md}.

\section{Scalings of radial disk profile and disk mass}\label{a:mass_scaling}

In this appendix we discuss the origin of the scalings observed in Figs. \ref{fig:scatter_Ms_Md} and \ref{fig:radial_profile}.
We begin by considering the scalings of $\Sigma$ and $T_{\rm mid}$. Approximately, our physical constraints are
\begin{eqnarray}
\Sigma &\propto& T^{1/2}\Omega \label{eq:Q_simple}\\
\Omega^2\dot M &\propto& \left\{\begin{array}{ll}T^4\Sigma^{-1}\kappa_{\rm R}^{-1}&{\rm (optically~thick)}\\T^4\Sigma\kappa_{\rm P}&{\rm (optically~thin)}\end{array}\right.\label{eq:thermal_eq_simple}
\end{eqnarray}
Here the first scaling corresponds to gravitational self-regulation ($Q\sim$constant) and the second corresponds to the thermal budget. $\kappa_{\rm R,P}$ are the average opacities at the given radius and $T$ is the average temperature (which should be comparable to the midplane temperature). This gives
\begin{eqnarray}
\Sigma &\propto& \left\{
\begin{array}{ll}
R^{-15/7}M_\star^{5/7}\dot M^{1/7} \kappa_{\rm R}^{1/7} & {\rm (optically~thick)}\\
R^{-5/3}M_\star^{5/9}\dot M^{1/9} \kappa_{\rm P}^{-1/9} & {\rm (optically~thin)}
\end{array}\right.
\label{eq:Sigma_scaling} \\
T &\propto& \left\{
\begin{array}{ll}
R^{-9/7}M_\star^{3/7}\dot M^{2/7} \kappa_{\rm R}^{2/7} & {\rm (optically~thick)}\\
R^{-1/3}M_\star^{1/9}\dot M^{2/9} \kappa_{\rm P}^{-2/9} & {\rm (optically~thin)}
\end{array}\right.
\end{eqnarray}
Here we have assumed that $M({<}R)\sim M_\star$, which is a reasonable order-of-magnitude approximation since we never have $M_{\rm d}/M_\star\gg 1$ in our sample.
The scalings of $\Sigma, T$ with respect to $R$ are consistent with those observed in Fig. \ref{fig:radial_profile}. (Here $\kappa_{R/P}$ should also vary in radius, but since $\Sigma$ and $T$ depend weakly on $\kappa_{R/P}$, we can ignore this dependence.)

Now we consider the scaling of disk mass. The disk mass is simply given by
\begin{equation}
M_{\rm d} = \int 2\pi \Sigma R^2 {\rm d}\log R.
\end{equation}
From Eq. \ref{eq:Sigma_scaling} we see that the $\Sigma R^2$ scales with $M_\star$ at a slope between $5/7=0.71$ and $5/9=0.56$, which is consistent with the empirical scaling of $M_{\rm d}\propto M_\star^{0.6}$. $\Sigma R^2$ is much less sensitive to other parameters (including $R$), and that explains why $M_{\rm d}$ shows such tight correlation with $M_\star$. Besides, the weak dependence of $\Sigma R^2$ on $R$ means that the relatively steep correlation between $M_{\rm d}$ and $R_{\rm d}$ in Fig. \ref{fig:scatter_Rd_Md} for larger disks is not just a generic feature of gravitationally self-regulated disks.

\end{CJK*}
\end{document}